\documentclass[10pt,aps,prmaterials,raggedbottom,longbibliography,nobalancelastpage,reprint,superscriptaddress,citeautoscript,letterpaper]{revtex4-2} 
\usepackage[usenames,dvipsnames]{color}
\usepackage{graphicx,microtype}
\usepackage[bookmarks=false,colorlinks]{hyperref}
\usepackage{lmodern}
\usepackage[all]{hypcap} 
\hypersetup{linkcolor=magenta,citecolor=MidnightBlue,filecolor=Plum,urlcolor=MidnightBlue}

\newcommand{\supf}{\textcolor{RoyalBlue}{Fig.}\,S}
\newcommand{\supt}{\textcolor{RoyalBlue}{Table}\,S}



\begin{document}

\title{Ferroelectricity in antiferromagnetic wurtzite nitrides}

\author{Steven M.\ Baksa}
\affiliation{Department of Materials Science and Engineering, Northwestern University, Evanston, Illinois  60208, USA}

\author{Lin-Ding Yuan}
\affiliation{Department of Materials Science and Engineering, Northwestern University, Evanston, Illinois  60208, USA}

\author{Stephen D.\ Wilson}
\affiliation{Materials Department, University of California, Santa Barbara, Santa Barbara, California 93106, USA}

\author{James M.\ Rondinelli}%
\email{jrondinelli@northwestern.edu}
\affiliation{Department of Materials Science and Engineering, Northwestern University, Evanston, Illinois  60208, USA}

\date{\today}
\begin{abstract}
 \noindent Wurtzite-type nitrides have recently emerged as promising candidates for ferroelectric applications, yet their magnetic counterparts remain largely unexplored. 
 Here, we establish MnSiN$_2$ and MnGeN$_2$ as aristotypes of a new multiferroic wurtzite family that simultaneously exhibits ferroelectricity and antiferromagnetism. 
 These Mn(II)-based nitrides crystallize in polar structures and display robust G-type antiferromagnetism at room temperature. 
 First-principles calculations reveal that nonmagnetic analogs incorporating Zn and Mg possess high polarization reversal barriers (0.735 and 0.683\,eV per formula unit) and wide band gaps (4.0 and 4.8\,eV), making them ideal ferroelectric candidates. 
 In contrast, MnSiN$_2$ and MnGeN$_2$ exhibit strong antiferromagnetic exchange interactions (5--9 meV per Mn site) and moderate band gaps (1.6 and 1.0 eV), with reversal barriers of 0.963 and 0.460 eV per formula unit, respectively. 
 Despite their limited magnetoelectric coupling, we show this family of Type-1 multiferroics exhibits altermagnetic spin splitting
 which reverses sign upon polarization switching.
 By strategically substituting alkaline-earth metals, we engineer multiple materials with coexisting switchable polarization, spin texture, and magnetic order. 
 These findings open new avenues for the design of nitride-based altermagnetic multiferroics, offering a platform for integrated antiferromagnetic spintronic devices. \\ \\
 \noindent \textbf{Keywords:} Ferroelectrics,  antiferromagnets, altermagnets, wurtzites, nitrides, spintronics
\end{abstract}

\maketitle

\section{Introduction}

Spintronic solid-state devices integrate nonequilibrium electron spins from an electric field or spin-polarized current \cite{Zutic2004, Wolf2006, Bader2010, Bhatti2017}, giving rise to functional electrical and optical phenomena \cite{Fukami2014, Fusil2014, Sinova2015, Coileain2017, Ling2017}. 
Recent research has explored antiferromagnetic (AFM) compounds for spintronic applications  \cite{Jungwirth2016, Gomonay2017,RevModPhys.90.015005,2025NatRM..10..109R}. 
AFM materials exhibit advantages, including stronger resistance to perturbations from magnetic fields than ferromagnetic (FM) materials, due to zero net magnetization, leading to stable data retention in storage applications, and quick switching times for read/write processes on the order of picoseconds in comparison to nanoseconds in FM spintronics. 
Despite these advances in high-density, energy-efficient spin-based electronic devices, most conventional AFMs are limited as many intrinsically 
exhibit spin degeneracy, inhibiting spin polarization. 
While recent advances have addressed this challenge through collinear altermagnets, noncollinear AFMs, and two-dimensional AFMs \cite{Guo2025}, one persistent challenge is the limited number of materials exhibiting AFM order at room temperature with significant electro-optical-magnetic coupling.

\begin{figure*}
    \centering
    \includegraphics[width=0.98\linewidth]{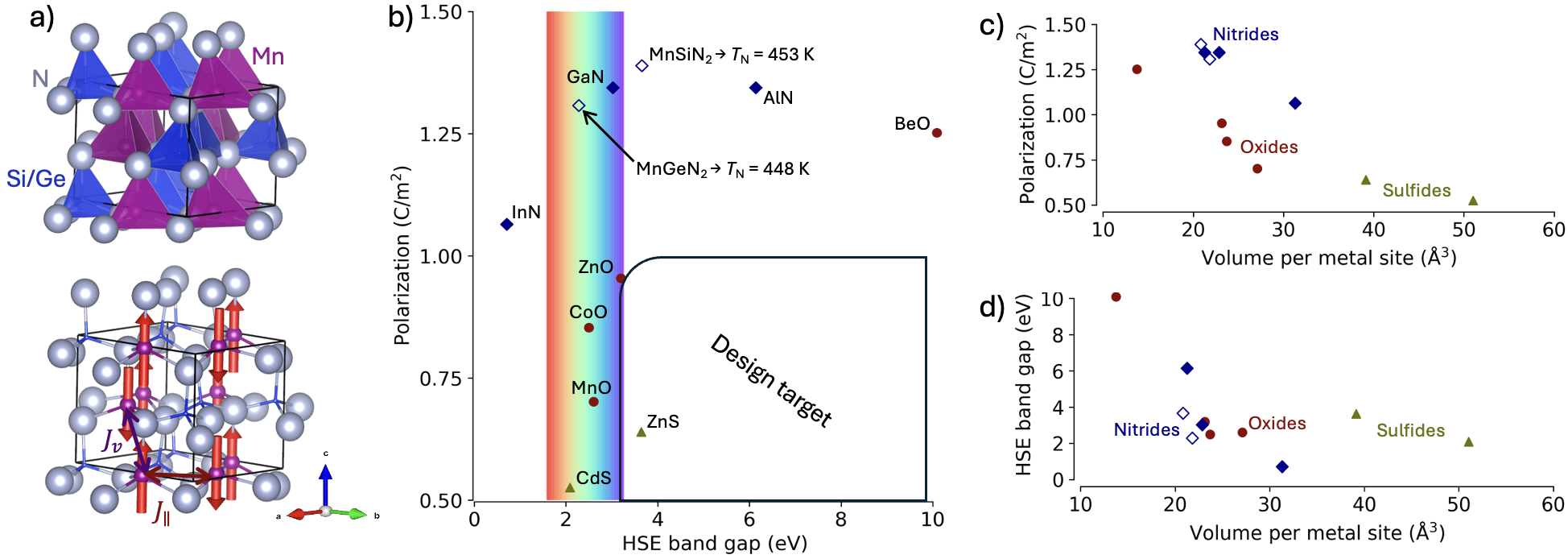}
    \caption{(a) Unit cell of wurtzite-type structure of Mn(Si,Ge)N$_2$ where Mn atoms are purple, Si/Ge atoms are blue, and N atoms are light gray. The compounds adopt G-type AFM (G-AFM) order and exhibit 
    two types of magnetic exchange:  diagonal inter-layer interaction, $J_{\rm v}$ (purple), and the in-plane intra-layer interaction, $J_\parallel$ (maroon). (b) Computed electric polarization as a function of the band gap computed at the DFT-HSE06 level. 
    from a representative sample of wurtzite oxides (red), sulfides (yellow), and nitrides (blue). 
    Magnetic compounds are denoted by open symbols. 
    A significant gap exists in the materials space for design targets with large optically insulating windows (\emph{i.e.,}, band gap $\ge$ 3.0 eV) and moderate electric polarization ($\le$ 1.00 C/m$^2$) for ferroelectricity. 
    (c) The electric polarization and (d) band gap as a function of the volume per metal site illustrate the challenge of optimizing both conditions in the search for magnetic ferroelectric wurtzites.}
    \label{fig:FE-AFM-setup}
\end{figure*}

One appealing class of AFM materials to pursue is wurtzite-type nitride compounds containing magnetic Mn$^{2+}$ cations ($d^5$ electronic configuration). 
These structures typically contain Group-IV elements like Si \cite{Esmaeilzadeh2006, Hausler2018, Kautzsch2023, Yuan2024}, Ge \cite{Hausler2018, Yuan2024, Lazarov2004}, and Sn \cite{Yuan2024, Rom2023} and exhibit $Pna2_1$ symmetry (\autoref{fig:FE-AFM-setup}a). 
MnSiN$_2$ and MnGeN$_2$ are G-type antiferromagnets (G-AFM) at room temperature with N\'eel temperatures ($T_{\rm N}$) of 453 K and 448 K, respectively \cite{Hausler2018, Kautzsch2023, Niewa2002}. 
These compounds are semiconducting with experimentally measured direct optical band gaps of 3.5 eV, 2.5 eV, and 1.2 eV for MnSiN$_2$ \cite{Hausler2018}, MnGeN$_2$ \cite{Hausler2018}, and MnSnN$_2$ \cite{Rom2023}, respectively.
In addition to the AFM order, these compounds are polar, leading to potential ferroelectricity, which has not previously been reported. 
Wurtzite crystal structures, including oxides, sulfides, and nitrides, have been proposed and have been demonstrated to exhibit ferroelectricity \cite{Calderon2023, wurtzite-fluorite2023, Wang2023}. 
To illustrate the challenges of finding optimal materials candidates simultaneously exhibiting ferroelectricity and AFM order, we have curated data comprising the electric polarization, band gap, and volume per metal site from a representative sample size 
(\autoref{fig:FE-AFM-setup}b). 
This representative sample of wurtzites includes ZnO \cite{Opoku2017, Bashyal2018}, BeO \cite{Shi2014}, CoO \cite{Tan2019}, MnO \cite{Tan2019}, ZnS \cite{Opoku2017}, CdS \cite{Shi2014}, AlN \cite{Duan2012}, GaN \cite{Duan2012}, and InN \cite{Duan2015} juxtaposed against MnSiN$_2$ and MnGeN$_2$.
The band gaps are based on the HSE06 exchange-correlation functional \cite{HSE06}, while the electric polarizations are computed following the Modern Theory of Polarization 
\cite{Resta2007, Spaldin2012}. 
Ideal ferroelectrics (FEs) are optically insulating (\emph{i.e.,} band gap $\ge$ 3.0 eV) to prevent dielectric breakdown with moderate electric polarization ($\le$ 1.00 C/m$^2$) to switch the poled state. 
Few wurtzite candidates satisfy both conditions since the electric polarization and band gap decrease with increasing volume (\autoref{fig:FE-AFM-setup}c,d). 
(Note that magnetic factors such as magnetic exchange or non-degenerate spin splitting are not accounted for, further limiting the materials design space.) 
To that end, ternary Si and Ge wurtzite-type nitrides containing divalent cations such as Zn \cite{Endo1992, Hausler2017, Ogura2021}, Mg \cite{Bruls1999, Quirk2014}, and Cd  \cite{Lyu2017} have been studied for their semiconducting properties with reported band gaps of 0.6 to 6.2 eV. %
Simulated electric polarization reversal has even been reported for MgSiN$_2$ and MgGeN$_2$ \cite{Lee2024}. 
Moreover, centrosymmetric manganese nitrides with $Pnma$ symmetry containing Ti, Zr, and Hf have been reported in the Materials Project database in a five-coordinated trigonal bipyramidal coordination environment \cite{MaterialsProject}. 
Lastly, it has been recently proposed that substitution of strongly electronegative cations like Si and Ge with less electronegative cations like Ti, Zr, or Hf can reduce the polarization reversal barrier in materials like MgSiN$_2$ \cite{Lee2024, Song2025}, opening opportunities for FE polarization reversal in derivative compounds.
Here, we design ferroelectric magnetic wurtzite-type nitrides \textit{via} targeted cation substitution into AFM semiconducting hosts to enhance their band gaps and enable multiferroic behavior. 
We propose the wurtzite-type $AB$N$_2$ series ($A$ = Mn, Zn, Mg, Cd, Ca; $B$ = Si, Ge, Ti, Zr, Hf) as a potential new family of materials exhibiting coupled FE and AFM properties. 
We systematically analyze the FE and AFM characteristics of the $A$(Si/Ge)N$_2$ end members, as well as ordered Mn(Si/Ge)N$_2$ compounds with 25\% molar substitution at the $A$- and $B$-sites.
Our results show that coexisting ferroelectricity and antiferromagnetism can be achieved by incorporating nonmagnetic cations into G-type AFM MnSiN$_2$ and MnGeN$_2$, offering promising pathways for the discovery and optimization of room-temperature multiferroics.
Lastly, we show that the altermagnetic spin splitting reverses sign upon ferroelectric switching. This behavior makes magnetic wurtzite nitrides compelling candidates for room-temperature electric field control of altermagnetic spin splitting \cite{PhysRevLett.134.106802,urru2025g}.

\section{Results and Discussion}
The end-member compounds we examine, in addition to G-AFM MnSiN$_2$ and MnGeN$_2$, include the orthorhombic wurtzite-type $A$SiN$_2$ and $A$GeN$_2$ ($A$ = Zn,Mg,Cd) with $Pna2_1$ symmetry, the tetragonal zincblende-type CaGeN$_2$ with $I\bar{4}2d$ symmetry, and the orthorhombic centrosymmetric Mn$B$N$_2$ ($B$ = Ti, Zr, Hf) with $Pnma$ symmetry [\supf1 and \supt1 of the Supporting Information (SI)]. 
While the $A^{2+}$-site and $B^{2+}$-site cations are tetrahedrally coordinated with N$^{3-}$ anions in the wurtzite-type and zincblende-type structures, these cations are in a five-coordinated trigonal bipyramidal environment with the N$^{-3}$ anions. 
The $A$-N bond lengths of both the wurtzite-type $A$SiN$_2$ and $A$GeN$_2$ series range from 1.9 to 2.3 \AA. 
The Si-N and Ge-N bond lengths are centered at 1.76 \AA\ and 1.90 \AA, respectively. 

Although we focus on the FE-AFM properties of ordered Mn(Si/Ge)N$_2$ compounds with 25\% molar substitution at the $A$- and $B$-sites, \emph{i.e.,} Mn$_3$A(Si/Ge)$_4$N$_8$ or Mn$_4$(Si/Ge)$_3B$N$_8$ compositions, we anticipate that similar behaviors occur in lower concentrations. 
More dilute systems are better described as wurtzite-type solid solutions, analogous to Sc- and B-doped AlN, where the substitutional species occupy uncorrelated lattice sites, yet still exert significant influence on the electronic structure, magnetic ordering, and ferroelectric switching characteristics.

    \subsection{Electronic structure}
\begin{figure}
        \centering
        \includegraphics[width=0.98\linewidth]{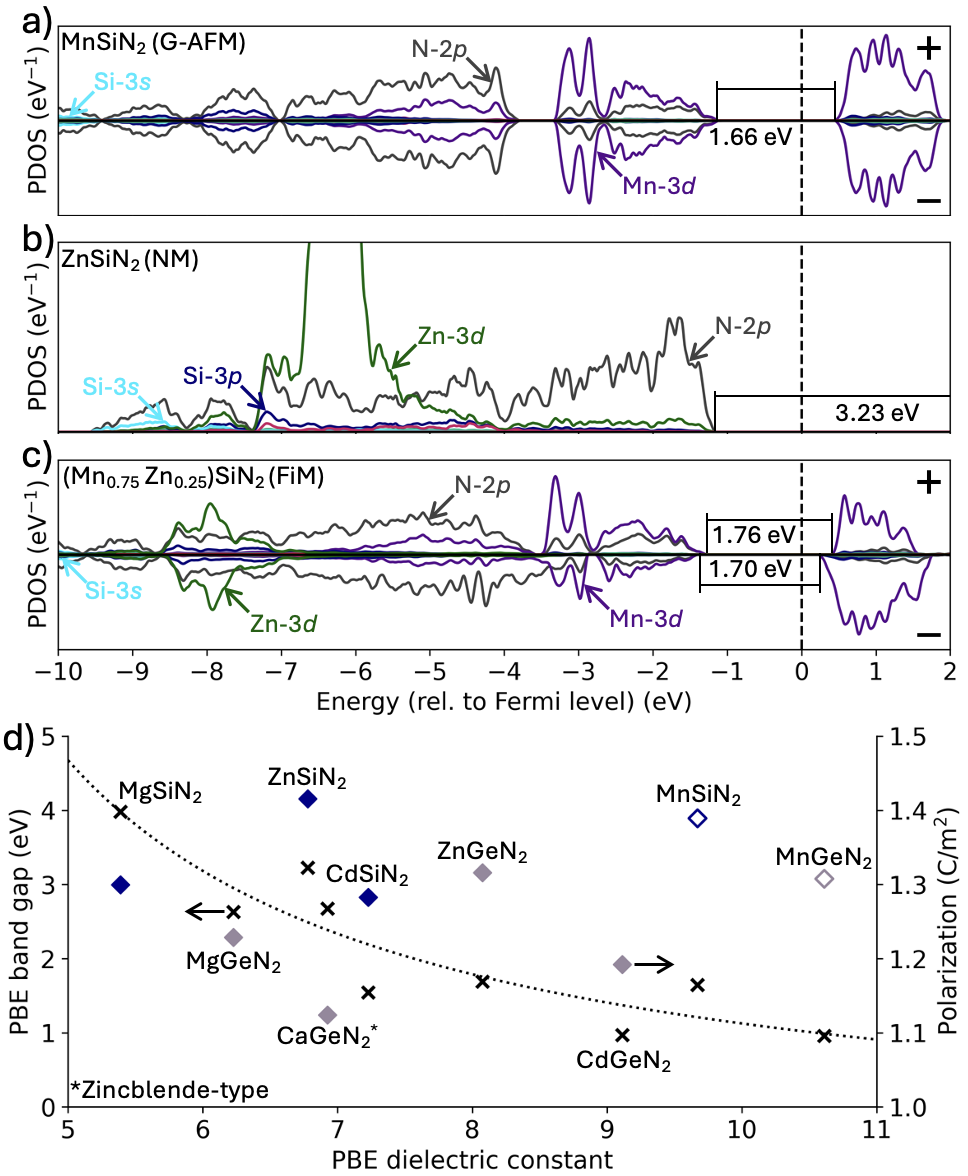}
        \caption{Projected density of states (PDOS) for (a) G-AFM MnSiN$_2$, (b) NM ZnSiN$_2$, and (c) FiM (Mn$_{0.75}$Zn$_{0.25}$)SiN$_2$. Orbital contributions: Si-3$s$ (sky blue), Si-3$p$ (dark blue), N-2$p$ (dark grey), Mn-3$d$ (purple), Zn-3$d$ (green), Zn-4$s$ (magenta). Zn substitution slightly increases the band gap with minimal dependence on magnetic state. (d) Band gaps of $A$SiN$_2$ and $A$GeN$_2$ (black crosses) versus electronic dielectric constant, consistent with the inverse gap–dielectric trend \cite{Penn1962, Wang2018}. Electric polarization (diamonds) shows no clear correlation with dielectric constant.}
        \label{fig:PDOS-dielectric}
    \end{figure}

    All compounds within the $A$(Si/Ge)N$_2$ series consist of MnN$_4$ tetrahedra (or more generally, divalent cation-nitride units) that are corner-shared with tetravalent Si/GeN$_4$ (or $B$N$_4$) tetrahedra;  these motifs exhibit strong $sp^3$ (or $dp$) hybridization, characteristic of covalently bonded tetrahedral networks.
    They are also valence-precise compounds, such that the frontier orbitals that form the band edges are primarily derived from the $A$-site cation states and nitrogen states, while the $B$-site cation states reside deeper within the valence band.
    To explore the electronic structure and dielectric response of Mn-based wurtzite-type nitrides and their respective ordered compounds, we show the atom- and orbital-resolved projected density of states (PDOS) of MnSiN$_2$, ZnSiN$_2$, and Mn$_{0.75}$Zn$_{0.25}$SiN$_2$ in \autoref{fig:PDOS-dielectric}a--c. 
    %
    Complete band structures, PDOS, dielectric function, and derived properties (\emph{e.g.,} refractive index, extinction coefficient), as well as the electro-optical (EO) coefficients, of all compounds are reported in \supf2-S7 and \supt2 of the SI. 
    Notable, the simulated EO coefficients are  an order of magnitude larger than those of main-group-based nitrides like (Al,Sc)N \cite{Pockels-Yang2024} and comparable to LiNbO$_3$ \cite{Wemple}.

    \begin{figure*}
        \centering
        \includegraphics[width=0.98\linewidth]{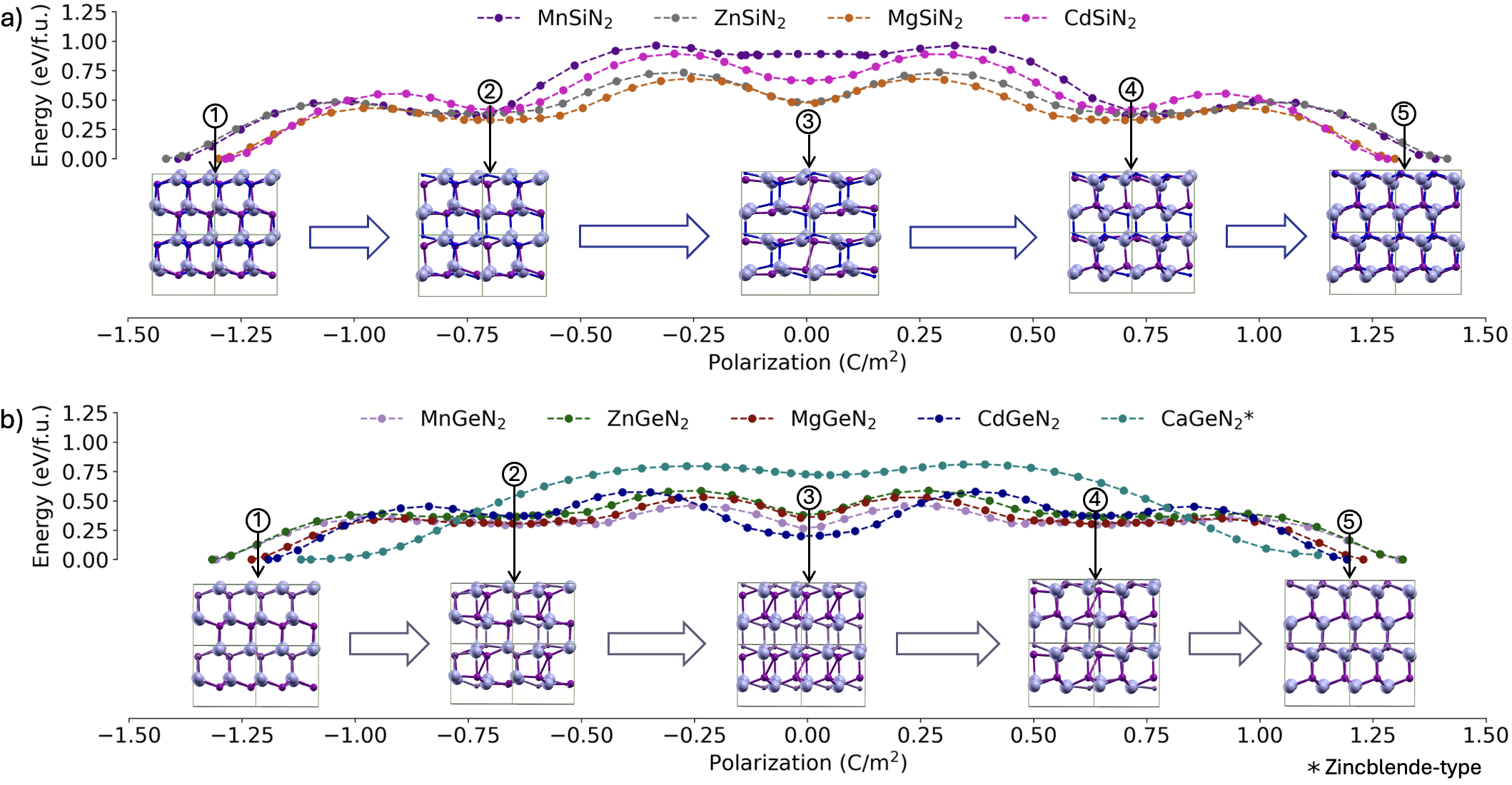}
        \caption{Minimum energy pathways of sequential FE polarization reversal in (a) $A$SiN$_2$ and (b) $A$GeN$_2$ end members, with intermediate images labeled 1-5 for clarity.  While MnSiN$_2$ passes through a significantly distorted non-polar phase with the largest reversal barrier of the $A$SiN$_2$ structures,  MnGeN$_2$ passes through an anti-polar phase, similar to substituted wurtzite nitrides like (Al,B)N \cite{Calderon2023}, with the smallest reversal barrier of the $A$GeN$_2$ structures. Note that CaGeN$_2$ is a zincblende-type structure, leading to a dissimilar reversal profile compared to the wurtzite-type $A$GeN$_2$ series.}
        \label{fig:NEB-pathways}
    \end{figure*}

    The Mn-3$d$ and N-2$p$ orbitals dominate the band edges of MnSiN$_2$, whereas the valence band edge of ZnSiN$_2$ is dominated by the N-2$p$ states with reduced contributions from the Zn-4$s$ orbitals in the conduction band edge. 
    Although the spin-polarized PDOS of MnSiN$_2$ is symmetric, it exhibits spin-splitting away from the $\Gamma$ point, making it altermagnetic \cite{Yuan2024}; additional spin-splitting can be induced at the $\Gamma$ point upon chemical substitution (\emph{vide infra}).
    %
    The computed band gaps of AFM MnSiN$_2$ and nonmagnetic (NM) ZnSiN$_2$ are 1.66 and 3.23 eV, respectively.  
    Substituting Zn into MnSiN$_2$ leads to a ferrimagnetic (FiM) ordered compound with exchange spin splitting in the PDOS. 
    Here, the band edges are dominated by the Mn-3$d$ and N-2$p$ orbitals with little contribution from the Zn-based orbitals, as expected. 
    The band gaps of majority-spin-up and majority-spin-down FiM Mn$_{0.75}$Zn$_{0.25}$SiN$_2$ are 1.76 and 1.70 eV, respectively. 
    By substituting with a NM cation like Zn, a slight enhancement of the band gap occurs relative to MnSiN$_2$ with little variance between the spin states. 
    MnSiN$_2$ is near the target band gap window suitable for advanced electronics, while ZnSiN$_2$ \cite{Hausler2017, Ogura2021} and MgSiN$_2$ \cite{Hausler2018, Fang1999} are strongly insulating with experimental band gaps of approximately 4.0 and 4.8 eV, respectively. 
    Although CdSiN$_2$, as well as CdGeN$_2$, has not been synthesized experimentally, its computed band gap decreases to 1.54 eV compared to 1.64 eV for MnSiN$_2$. 
    MnGeN$_2$, on the other hand, has the lowest experimental band gap at 2.5 eV \cite{Hausler2018}, whereas ZnGeN$_2$ \cite{Hausler2017, Ogura2021} and MgGeN$_2$ \cite{Hausler2018} have improved insulating behavior, with each having band gaps at approximately 3.2 eV. 
    The computed band gap of CdGeN$_2$ is less than 1 eV. 
    All centrosymmetric Mn$B$N$_2$ structures are found to be metallic within our DFT-PBE calculations (\supf3-4 of the SI).
    %
    These electronic structures indicate that MnSiN$_2$, ZnSiN$_2$, MgSiN$_2$, ZnGeN$_2$, and MgGeN$_2$ exhibit appealing electronic properties; although the band gap of MnGeN$_2$ falls below the design target, cation substitution could drive insulating behavior in the ordered compound.
    The electronic dielectric constant, plotted against the band gap at the DFT-PBE level (\autoref{fig:PDOS-dielectric}d), reveals an inverse correlation in agreement with theoretical predictions and experimental observations across diverse material systems \cite{Wang2018}.
    One way to rationalize this relationship is to consider charge carriers, which have an Arrhenius relationship with the band gap. 
    Since the dielectric constant is a measure of the ability to polarize charge carriers, it also exhibits this Arrhenius behavior. 
    The magnetic end members have the lowest band gaps (largest dielectric constants), which derives from extensive orbital overlap due to the strong $\sigma$ and $\pi$ bonding character between the nitrogen anion and the transition metal cation \cite{Kautzsch2023}.
    The NM compounds, in contrast, tend to have larger band gaps (lower dielectric constants), with the silicon nitrides ($A$SiN$_2$--blue diamonds) having the largest band gaps compared to the germanium nitrides due to bond covalency. 
    The computed electric polarization of $A$SiN$_2$ (blue diamond) and $A$GeN$_2$ (gray diamond), including G-AFM MnSiN$_2$ and MnGeN$_2$ (open diamond) shows no clear trend with respect to the dielectric constant.

    \subsection{Ferroelectric polarization reversal}

    \subsubsection{Pristine ternary nitrides}
    While a uniform switching model is applied to bulk binary wurtzites \cite{Moriwake2014, Moriwake2020, Yang2024}, the primary mechanism in substituted and complex wurtzites is columnar switching, where metal-nitrogen columns oriented along the $\langle001\rangle$ crystallographic direction are switched sequentially until all columns are switched. 
    \autoref{fig:NEB-pathways} shows the computed minimum energy pathways in the $A$SiN$_2$ and $A$GeN$_2$ series.
    %
    The $A$SiN$_2$ compounds pass through a non-polar phase with significant distortions, that is, one that significantly deviates from the $h$-BN intermediate phase commonly reported in uniform FE reversal pathways of wurtzites. 
    The $A$GeN$_2$ compounds pass through an anti-polar phase that is reminiscent of FE reversal pathways of substituted wurtzite nitrides such as (Al,B)N \cite{Calderon2023}. 
    Moreover, all FE reversal pathways are sequential compared to uniform switching computed in bulk binary wurtzites \cite{Moriwake2014, Moriwake2020}. 
    Monitoring the transition from $-P$ to $+P$ states, we find that the $A$SiN$_2$ and $A$GeN$_2$ compounds transition from $-P$ (1) to $+P$ (5) in eight sequential steps, where each hill of the minimum energy pathway contains one step at the peak and one step at the valley (labeled 2-4).
    Although the zincblende-type CaGeN$_2$ passes through an anti-polar phase, the sequential switching is restricted to four steps rather than eight steps observed in the wurtzite-type series. 
    This key difference can be rationalized based on stronger distortions in the CaGeN$_2$ in comparison to the wurtzite-type end members.
    The significance of these pathways is evident by comparing the reversal barriers $\Delta$ of MnSiN$_2$ (0.963 eV/f.u.) and MnGeN$_2$ (0.460 eV/f.u.) with those of the $A$-substituted Si and Ge nitride compounds (\autoref{tab:data}).
    %
    %
    The electric polarization of all ternary compounds ranges from 1.2 to 1.4 C/m$^2$ (\autoref{fig:NEB-pathways}); full details of the FE and related properties of all wurtzite- and zincblende-type end members are reported in \supf11 of the SI. 
    In general, the $A$GeN$_2$ series exhibits lower reversal barriers and electric polarization values than the $A$SiN$_2$ series, attributed to the different sequential reversal pathways, cation radii, and bond ionicity (see SI). 
    Among the $A$GeN$_2$ compounds, zincblende-type CaGeN$_2$ shows the largest $\Delta=0.81$\,eV/f.u.\ with a different FE reversal profile compared to the wurtzite-type members. 
    Based on the minimum energy pathways, electric polarizations, and electronic structures, NM ZnSiN$_2$ and MgSiN$_2$ emerge as the strongest candidates for FE polarization reversal in the $A$SiN$_2$ series due to their low reversal barrier and high band gap. 
    MnSiN$_2$ may also be viable if its electronic band gap can be increased and the reversal barrier reduced (\emph{vide infra}). 
    For the $A$GeN$_2$ series, NM ZnGeN$_2$ and MgGeN$_2$ are the strongest candidates.
    %
    Although MnGeN$_2$ has the lowest reversal barrier among the ternary wurtzite-type compounds considered, its band gap falls below the design window to prevent dielectric breakdown.
    To address this, we investigate cation substitution in MnGeN$_2$ to enhance its insulating behavior while retaining its low switching barrier. 

\begin{table}
\centering
\begin{ruledtabular}\
\caption{Summary of ferroelectric polarization reversal barriers $\Delta$ (eV/f.u.) and mixing enthalpy $\Delta H_{\rm mix}$ (meV/f.u.) of $A$-site Si and Ge nitride ordered compounds with 25\% cation substitution. The four-coordinate Shannon-Prewitt radii ($r_\mathrm{SP}$) are provided in \AA\ \cite{Shannon1976}; the corresponding four-coordinate radii for Si$^{4+}$ and Ge$^{4+}$ are 0.26\,\AA\ and 0.39\,\AA, respectively.}
\begin{tabular}{cccccc}
             & & \multicolumn{2}{c}{Si}                                & \multicolumn{2}{c}{Ge}                                 \\ 
             \cline{3-4}\cline{5-6}
$A^{2+}$ & $r_\mathrm{SP}$ & \multicolumn{1}{c}{$\Delta$} & \multicolumn{1}{c}{$\Delta H_{\rm mix}$} & \multicolumn{1}{c}{$\Delta$} & \multicolumn{1}{c}{$\Delta H_{\rm mix}$} \\
\hline
Mn$^{2+}$ & 0.66    & 0.963                       & -                        & 0.46                        & -                        \\
Mg$^{2+}$ &  0.57    & 0.683                       & 44.3                     & 0.533                       & 33.6                     \\
Zn$^{2+}$ & 0.60    & 0.735                       & 57.4                     & 0.588                       & 44.4                     \\ 
Cd$^{2+}$ & 0.78    & 0.893                       & 94.7                     & 0.578                       & 66.1                    
\end{tabular}
\end{ruledtabular}
\label{tab:data}
\end{table}

    \subsubsection{Cation-ordered nitrides}
    Ordered compounds of MnSiN$_2$ and MnGeN$_2$ \textit{via} $A$-site and $B$-site cation substitution at 25\% dopant molar concentration are considered based on the appeal of the bridging ferroelectricity and antiferromagnetism in a single materials system. Factors that contribute to the stability and formation of ternary and quaternary nitrides include electronegativity, ionic size, and charge effects \cite{Won2025}. The thermodynamic feasibility of 25\% cation substitution in MnSiN$_2$ and MnGeN$_2$ was assessed by computing the mixing enthalpy of each ordered compound as %
    $    \Delta H_{\rm mix} = H_{\rm ordered} - (1-x)H_{\rm C_1} -xH_{C_2}$, 
    where $H_{\rm ordered}$, $H_{\rm C_1}$, and $H_{\rm C_2}$ are the enthalpies of the ordered compound, $C_1$, and $C_2$, respectively, and $x$ is the molar fraction of end member $C_2$ ($x=0.25$). 
    The $\Delta H_{\rm mix}$ of the $A$-site Si-based ordered compounds is generally more positive than the $A$-site Ge-based ordered compounds (\autoref{tab:data}), whereas the $\Delta H_{\rm mix}$ of $A$-site ordered compounds is exceedingly less positive than the $B$-site ordered compounds. 
    The $\Delta H_{\rm mix}$ of the divalent $A$-site ordered compounds is less positive than the tetravalent $B$-site ordered compounds. 
    The $B$-site Si-based ordered compounds in increasing $\Delta H_{\rm mix}$ are Mn(Si,Ti)N$_2$ (139.6 meV/f.u.), Mn(Si,Hf)N$_2$ (151.1 meV/f.u.), and Mn(Si,Zr)N$_2$ (170.2 meV/f.u.), whereas the $B$-site Ge-based ordered compounds in increasing $\Delta H_{\rm mix}$ are Mn(Ge,Hf)N$_2$ (98.1 meV/f.u.), Mn(Ge,Zr)N$_2$ (112.7 meV/f.u.), and Mn(Ge,Ti)N$_2$ (114.8 meV/f.u.). 
    %
    These mixing enthalpies indicate that $A$-site Ge-based ordered compounds, as well as $A$-site Si-based ordered compounds, to a lesser extent, are generally the most promising candidates to pursue based 
    on thermodynamic metastability considerations.

     \begin{figure}
        \centering
        \includegraphics[width=0.98\linewidth]{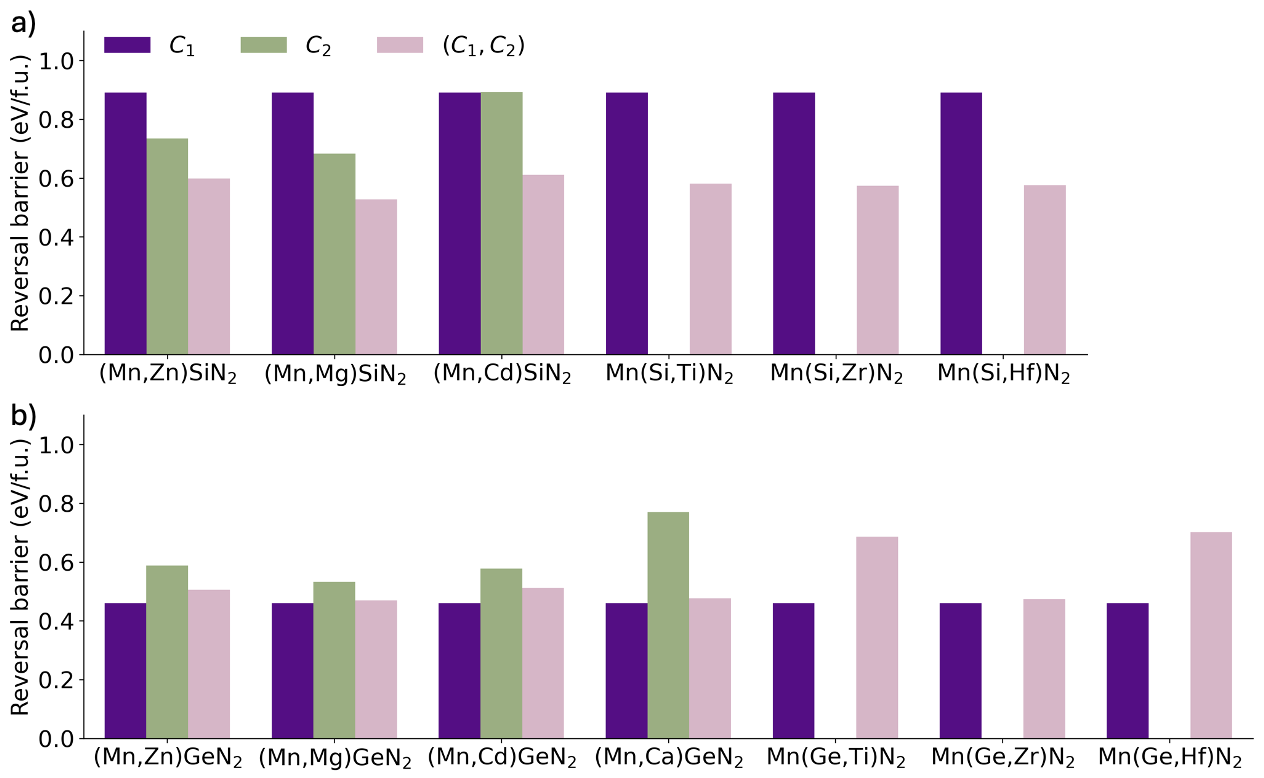}
        \caption{FE polarization reversal barriers of end members and ordered (a) Si-based and (b) Ge-based nitrides with Mn-based nitrides ($C_1$) in purple, NM and non-Si/Ge nitrides ($C_2$) in green, and ordered compounds at 25\% molar substitution [($C_1$, $C_2$)] in pink. End members containing Ti, Zr, or Hf are centrosymmetric with zero reversal barrier. While ordered Si-based nitrides have a lower reversal barrier than their polar end members, ordered Ge-based nitrides have a marginally larger reversal barrier than MnGeN$_2$ except for Mn(Ge,Ti)N$_2$ and Mn(Ge,Hf)N$_2$. }
        \label{fig:Doped-energy-barriers}
    \end{figure}

    We then compute the FE reversal barriers of $A$-site and $B$-site cation-ordered nitrides of MnSiN$_2$ and MnGeN$_2$ at 25\% molar concentration (\autoref{fig:Doped-energy-barriers}). 
    Here, $C_1$ is in purple, $C_2$ is in green, and the ordered compound [($C_1$, $C_2$)] is in pink. 
    Recall that the Mn$B$N$_2$ series is centrosymmetric with zero reversal barrier. 
    $A$-site cation substitution resulted in a lower reversal barrier than either end-member structure for the Si-based ordered compounds and a marginal increase in the reversal barrier for the Ge-based ordered compounds. 
    Recall that the intermediate non-polar phase of MnSiN$_2$ contains significant distortions, whereas the intermediate anti-polar phase of MnGeN$_2$ contains minimal distortions. 
    Thus, distortions from cation substitution in MnSiN$_2$ may drive the system to engage in symmetry breaking, thereby finding a smaller reversal barrier, whereas similar cation substitution in MnGeN$_2$ may not achieve the same effect. 
    With $B$-site cation substitution, we find smaller and larger reversal barriers in the Si-based and Ge-based ordered compounds, respectively, except Mn(Ge,Zr)N$_2$, which only showed a marginal increase in the reversal barrier. 
    This exception can be rationalized by considering thermodynamic stability. 
    While the $\Delta H_{\rm mix}$ of the $B$-site Si-based ordered compounds increases linearly from Ti to Hf, the $\Delta H_{\rm mix}$ of the $B$-site Ge-based ordered compounds plateaus between Zr and Hf. 
    We note that the MnGeN$_2$ end member exhibited the lowest reversal barrier out of the end members and the ordered compounds. 

        \begin{figure*}
        \centering
        \includegraphics[width=0.98\linewidth]{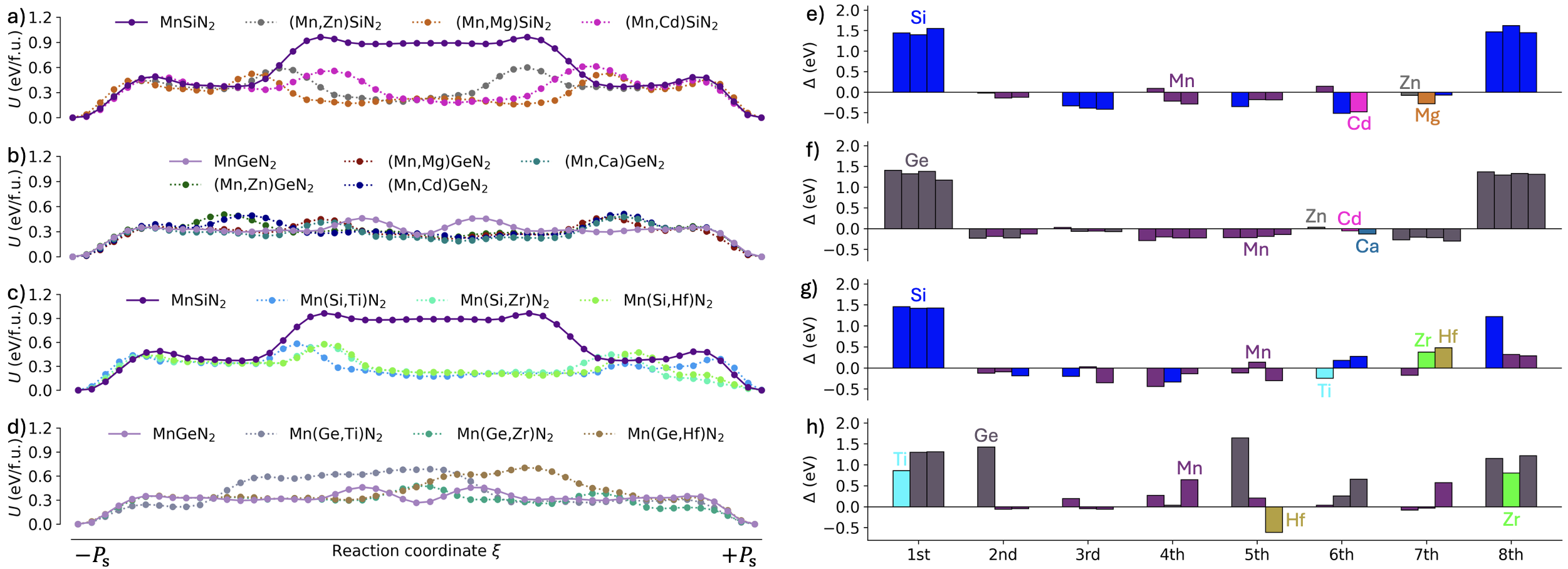}
        \caption{Minimum energy pathways of the (a) (Mn,$A$)SiN$_2$, (b) (Mn,$A$)GeN$_2$, (c) Mn(Si,$B$)N$_2$, and (d) Mn(Ge,$B$)N$_2$ series computed \textit{via} nudged elastic band with the reaction coordinate $\xi$ from negative to positive polarity ($-P_{\rm s} \rightarrow +P_{\rm s}$). The MnSiN$_2$ and MnGeN$_2$ end members are shown as solid lines for clarity. Most ordered compounds exhibit flat regions in the minimum energy pathway, where the energy remains constant as a function of $\xi$. The atomic switching barrier $\Delta$ from the 1st to 8th metal-nitrogen pair for the (e) (Mn,$A$)SiN$_2$, (f) (Mn,$A$)GeN$_2$, (g) Mn(Si,$B$)N$_2$, and (h) Mn(Ge,$B$)N$_2$ series. The series for the 1st to 8th cation is organized based on the legend order from the corresponding minimum energy pathways. The Mn (purple), Zn (gray), Mg (orange), Cd (pink), Ca (teal), Si (royal blue), Ge (periwinkle), Ti (sky blue), Zr (green), and Hf (chartreuse) cations are labeled. While these results largely confirm that the most electronegative cation (\emph{i.e.,} Si, Ge) drives the FE polarization reversal with the largest $\Delta$, some $B$-type ordered compounds exhibit a nuanced switching order, including the Mn(Ge,$B$)N$_2$ series with significant $\Delta$ values throughout the pathway.}
        \label{fig:NEB-atomic-probe}
    \end{figure*}

    To better understand the atomistic origins of the driving mechanism of the FE polarization reversal barrier in the end members and their respective ordered compounds, we analyzed the minimum energy pathways as a function of the reaction coordinate $\xi$ (from $-P_{\rm s}$ to $+P_{\rm s}$) of the $A$SiN$_2$, $A$GeN$_2$, (Mn,$A$)SiN$_2$, (Mn,$A$)GeN$_2$, Mn(Si,$B$)N$_2$, and Mn(Ge,$B$)N$_2$ series 
    (\autoref{fig:NEB-atomic-probe}a-d and \supf9 of the SI). 
    The G-AFM MnSiN$_2$ and MnGeN$_2$ are shown in solid lines for clarity, and 
    are the same as those in \autoref{fig:Doped-energy-barriers}. 
    While some NM end members, including ZnSiN$_2$ and MgSiN$_2$, exhibit regions of substantial concavity in the minimum energy pathway, most ordered compounds exhibit flat regions in the minimum energy pathway, where the energy does not change with respect to $\xi$. 
    Two possibilities for this phenomenon include (1) translation of the crystal structure that can arise as an artifact of the nudged-elastic-band simulation or (2) little to no reversal barrier from individual metal-nitrogen pairs switching from negative to positive polarity. 
    To determine the cause of this behavior, the $\Delta$ of each metal-nitrogen pair
    in each series are computed as 
    $    \Delta_{i} = U_{i}-U_{i-1}$ (\autoref{fig:NEB-atomic-probe}e-h, \supf9 of the SI), 
    where $i$ represents the $i$-th atom switched from negative to positive polarity. 
    Here, positive $\Delta_i$ corresponds to an energy barrier to switching, while a negative $\Delta_i$ corresponds to barrierless switching. The series for the 1st to 8th cation is organized based on the legend order from the corresponding minimum energy pathways. 
    Our results indicate that flat regions in the minimum energy pathway are not an artifact, but stem from low to barrierless FE reversal of the metal-nitrogen pairs. 

    The $\Delta$ values of the $A$SiN$_2$ and $A$GeN$_2$ end members reveal key trends in the switching patterns (\supf9 of the SI). 
    Both series exhibit the highest $\Delta$ values for the 1st and 8th metal-nitrogen pair, corresponding to the  Si/Ge cations, \emph{i.e.,}  the most electronegative cations in the families, 
    consistent with previously proposed design rules for FE polarization reversal in the wurtzite-type materials \cite{Lee2024, Song2025}. 
    The Pauling electronegativity of the relevant cations are: Ge$^{4+}$ (2.01), Si$^{4+}$ (1.90), Cd$^{2+}$ (1.69), Zn$^{2+}$ (1.65), Ti$^{4+}$ (1.54), Mn$^{2+}$ (1.55), Zr$^{4+}$ (1.33), Mg$^{2+}$ (1.31), Hf$^{4+}$ (1.30), and Ca$^{2+}$ (1.00). 
    Substituting Si/Ge with less electronegative tetravalent cations, \emph{e.g.,} Ti, Zr, and Hf, has been proposed to reduce the reversal barrier in the NM ternary nitrides \cite{Lee2024}, which we further assess for the magnetic nitrides. 

    MnSiN$_2$ exhibits a Si-Mn-Si-Mn switching sequence with large $\Delta$ values for Si, while MnGeN$_2$ follows a Ge-Mn-Mn-Ge sequence with smaller $\Delta$ values for the intermediate Ge cations. 
    In comparison, NM $A$SiN$_2$ compounds show lower $\Delta$ values for both $A$-site cations and intermediate Si cations than G-AFM MnSiN$_2$, whereas the $\Delta$ values in G-AFM MnGeN$_2$ are similar to those in NM $A$GeN$_2$. 
    While the Zn, Mg, and Cd Si-based end members exhibit small $\Delta$ values, Cd in CdGeN$_2$ shows a significantly negative (barrierless) $\Delta$. 
    These comparisons underscore the value of analyzing individual cation contributions to understand and engineer FE polarization reversal barriers.
    
    We expand this analysis to $A$-site (\autoref{fig:NEB-atomic-probe}e,f) and $B$-site (\autoref{fig:NEB-atomic-probe}g, h) ordered compounds. 
    %
    In both cases, the Si/Ge cations consistently occupy the 1st and 8th positions with the largest $\Delta$ values, while all other cations exhibit either negligible or barrierless $\Delta$, accounting for the improved FE performance in the (Mn,$A$)SiN$_2$ series and the modest hindrance observed in the (Mn,$A$)GeN$_2$ series.
    For these ordered compounds, the primary challenge lies in switching the first metal-nitrogen pair; subsequent pairs switch with little to no resistance. 
    In contrast, the $B$-type ordered compounds exhibit a more complex and non-trivial switching behavior. 
    Although every ordered compound shows Si/Ge as switching in either the 1st or 8th position, Mn(Si,Zr)N$_2$ and  Mn(Si,Hf)N$_2$ show Mn switching in the 8th position whereas Mn(Ge,Ti)N$_2$ and Mn(Ge,Zr)N$_2$ have $B$-site cations in those positions. 
    The Mn(Si,$B$)N$_2$ series follows a trend consistent with the electronegative-cation design rule; however, this trend does not occur for the Mn(Ge,$B$)N$_2$ series. 
    There are small $\Delta$ values for intermediate cations in the Mn(Si,$B$)N$_2$ series, with Ti being barrierless and Zr and Hf demonstrating moderate $\Delta$ values. 
    The Mn(Ge,$B$)N$_2$ series, in contrast, exhibits significant $\Delta$ values for intermediate Mn and Ge cations, except for Mn(Ge,Zr)N$_2$ as evidenced by the marginal change in the overall FE polarization reversal barrier. 

    The atomistic analysis explains the substantial change in behavior between $A$-site and $B$-site ordered compounds, the similar performance between the (Mn,$A$)SiN$_2$ and Mn(Si,$B$)N$_2$ series, and the diverging trends between the Mn(Si,$B$)N$_2$ and Mn(Ge,$B$)N$_2$ series, while also resolving the anomaly found in Mn(Ge,Zr)N$_2$. 
    We find that it is insufficient to substitute strongly electronegative cations with less electronegative cations to reduce the FE polarization reversal barrier, which is why it is critical to account for defect interactions.
    Overall, $A$-site ordered Si-based compounds show the most promising reduction in reversal barriers, while Ge-based ordered compounds exhibit only marginal improvements. 
    Specifically, we recommend FE experiments on Zn/Mg-substituted ordered compounds of MnSiN$_2$, and to a lesser extent, MnGeN$_2$.

    \begin{figure}
        \centering
        \includegraphics[width=0.75\linewidth]{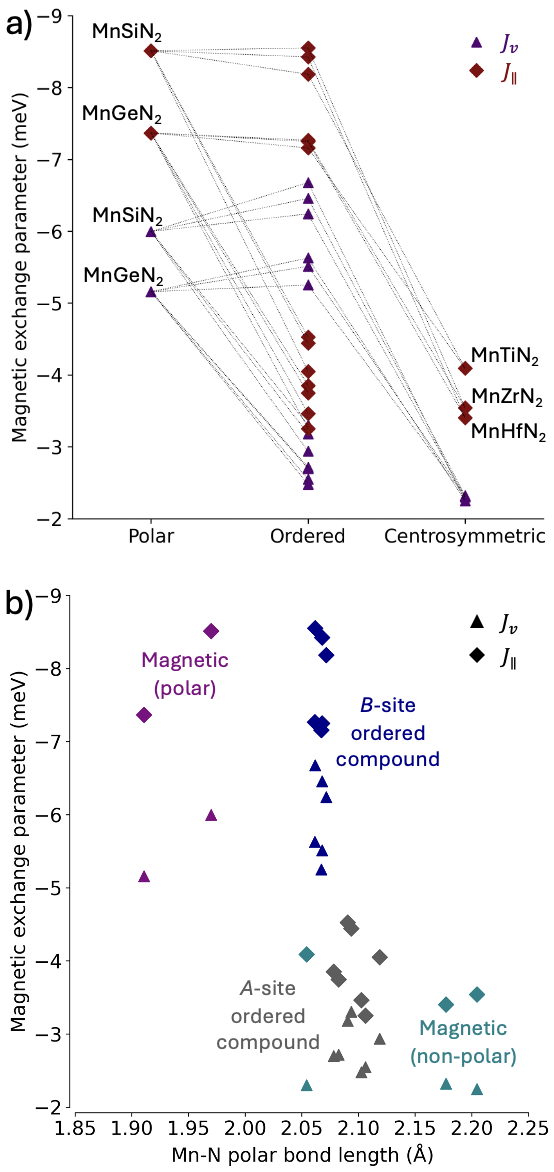}
        \caption{(a) Magnetic exchange parameters ($J_{\rm v}$ and $J_{\rm \parallel}$) for polar and centrosymmetric end members, and ordered compounds with 25\% substitution. 
        Dotted lines indicate ordered derivatives. 
        Despite reduced AFM exchange from NM substitution, G-AFM order persists in all ordered compounds, 
        as evidenced by $J<0$. 
        (b) Exchange parameters as a function of the Mn-N polar bond length. 
        Magnetic (polar) [purple], magnetic (non-polar) [teal], $A$-site ordered compound [gray], and $B$-site ordered compound [blue] cases are shown.
        Shorter Mn–N bonds generally correlate with stronger AFM exchange.}
        \label{fig:Magnetic-exchange}
    \end{figure}

    \subsection{Antiferromagnetic exchange}
    We compute the magnetic exchange parameters of the Mn-based nitrides and their respective ordered compounds using a simplified Heisenberg spin Hamiltonian (see SI). 
    The $J_{\rm v}$ (purple) parameter represents an interaction between the Mn planes, while the $J_{\rm \parallel}$ (maroon) parameter represents an in-plane interaction (\autoref{fig:FE-AFM-setup}a). 
    Our total energy calculations indicate that all magnetic end member and ordered compounds have G-AFM spin ordering in the ground state; other spin orders, including ferromagnetic (FM), A-type (A-AFM), and C-type (C-AFM), have higher energies (\supf8 of the SI). 
    Next, we examine the evolution in $J_{\rm v}$ and $J_{\rm \parallel}$ as MnSiN$_2$ and MnGeN$_2$ are modified through cation substitution to form ordered compounds (indicated with dotted lines in \autoref{fig:Magnetic-exchange}a). 
    The wurtzite-type magnetic end members consistently have stronger AFM interactions than the centrosymmetric counterparts. 
    This trend can be attributed to differences in the Mn-N polar bond lengths in the MnN$_4$ tetrahedra (\autoref{fig:Magnetic-exchange}b), where magnetic (polar) [purple], magnetic (non-polar) [teal], $A$-site ordered compound [gray], $B$-site ordered compound [blue] cases are illustrated. 
    The polar Mn-N bond lengths in the centrosymmetric (non-polar) end members are consistently larger than those of the polar end members and their ordered derivatives, resulting in reduced AFM exchange strength within the centrosymmetric structures.  
    Moreover, ordered compounds derived from magnetic end members (both wurtzite-type and centrosymmetric) have stronger AFM exchange than those incorporating  NM cations, due to disruption of the AFM exchange network upon $A$-site cation substitution. 
    Thus, while cation substitution is a powerful strategy for tuning ferroelectric properties, it can also enhance magnetic exchange interactions when applied deliberately, enabling the design of nitrides with optimal ferroelectric properties and robust magnetic ordering. 
    Although no clear trend emerges within individual subgroups, \emph{e.g.,} such as polar magnetic end members, the overall exchange dependence on Mn-N polar bond length reveals a sharp decline as it increases, consistent with previously reported trends \cite{Wang2021}.
    %

\begin{figure}
    \centering
    \includegraphics[width=\linewidth]{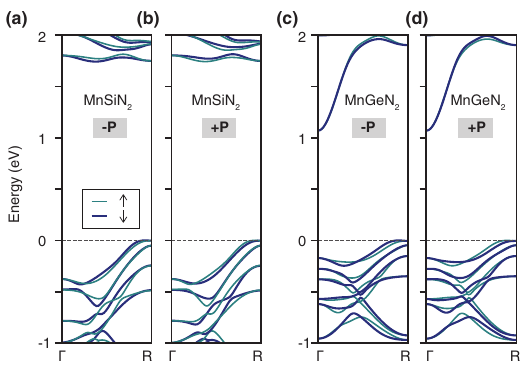}
    \caption{Switching of the nonrelativistic spin splitting (NRSS) in ferroelectric (a,b) MnSiN$_2$ and (c,d) MnGeN$_2$. Energy band dispersions of MnSiN$_2$ and MnGeN$_2$ in the (a,c) negatively  ($-P_\mathrm{s}$) and (b,d) positively ($+P_\mathrm{s}$) polarized state, respectively.}
    \label{fig:MnGeN2_nrss_bands}
\end{figure}
    
    \subsection{Magnetoelectric coupling and altermagnetism}
    Finally, we investigate magnetoelectric (ME) coupling in the form of changes to the AFM order during ferroelectric polarization reversal without spin-orbit coupling. 
    We compute the magnetic reversal barriers for MnSiN$_2$ and MnGeN$_2$ at fixed, poled structures under constrained magnetization across various magnetic configurations (FM, A-AFM, C-AFM, G-AFM), along with FE reversal barriers as a function of spin order (\supf12 of the SI).
    While the constrained magnetic reversal barriers are approximately 2.0 eV/f.u., the FE reversal barriers are reduced by a factor of two and four for MnSiN$_2$ and MnGeN$_2$, respectively. %
    The G-AFM order persists throughout the FE polarization reversal with minimal disruptions to the magnetic moment centered on the Mn$^{2+}$ ($d^5$) cations (\supf10 of the SI).
    %
    %
    Although MnSiN$_2$ and MnGeN$_2$ exhibit limited linear ME coupling, the robust AFM order in the presence of a perturbing electric field highlights the potential of this class of nitrides for spintronic applications.

Moreover, MnSiN$_2$ and MnGeN$_2$ are predicted to be altermagnets, exhibiting nonrelativistic spin splitting (NRSS) even without relativistic spin-orbit coupling. 
In $A$-site cation-ordered nitrides, large NRSS is demonstrated, even at the $\Gamma$ point, as a consequence of symmetry breaking \cite{Yuan2024}.
The NRSS arises from the coupling between the AFM ordering and the alternating local crystal environment due to the nitride anions; thus, the spin-structure motif pairs \cite{Yuan2023-AM} defining the positively and negatively poled states lock the spin-dipole orientation.  
Switching the local crystal environment would then reverse the AFM order and, therefore, reverse the splitting (\autoref{fig:MnGeN2_nrss_bands}). 
MnSiN$_2$ and MnGeN$_2$, along with their cation-ordered variants (see Section VII of the SI), provide a promising platform for realizing room-temperature electric-field switchable altermagnets.
Although several recent studies have demonstrated that inversion domains in wurtzite materials can exhibit structural and functional behavior analogous to 180° ferroelectric domains in conventional perovskites \cite{Lee2024,Skidmore2025,Fichtner2025,HuangArxiv}, we highlight that in the chemically-ordered wurtzite system, critical distinctions emerge when examining the momentum-dependent NRSS (\supf13).
In a wurtzite inversion domain, the positive and negative polarity orientations are related by a true crystallographic inversion operation, which is equivalent to $2/m$ symmetry, i.e., a $180^\circ$ rotation followed by a mirror reflection perpendicular to the rotation axis. 
This operation fully reverses the local site symmetry \textit{via} tetrahedral inversion, resulting in a complete reversal of the stacking sequence and crystal field environment. When these are valid operations of the crystal, opposite spin polarizations will exhibit reversed splittings. 
In contrast, a $180^\circ$ domain in a ferroelectric material involves a polarization rotation within a shared lattice framework such that the twins are related by a mirror. 
This is the switching path that we computationally modeled in \autoref{fig:NEB-pathways} and has been studied in wurtzites by others. 
The domain wall embodies the mirror symmetry element that reverses the direction of the polarization.
Therefore, these two orientational domains are expected to exhibit the same spin splitting as they are related by a mirror that does not switch the local crystal field environment on the spin sublattices. 
Chemical ordering with 25\% substitution on the $B$-site 
further lowers the symmetry, 
and renders the inversion domain and $180^\circ$ domain inequivalent.
The local crystal field environments are also different; this inequivalence appears in the nonrelativistic spin-splitting  for the two types of domains (\supf13). 
The inversion domain rather than the $180^\circ$ domain exhibits a reversal of the NRSS upon switching the polarization.
%
In summary, while both domain types may appear similar in terms of macroscopic polarization reversal, how one interconverts $-P_{\rm s} \leftrightarrow +P_{\rm s}$ upon switching produces fundamentally distinct NRSS responses that warrant experimental investigation.

\section{Conclusion}
We presented a comprehensive analysis of electronic, ferroelectric, antiferromagnetic (altermagnetic), and magnetoelectric (ME) properties of MnSiN$_2$, MnGeN$_2$, and their ordered derivatives. 
We discovered that the ternary Zn/Mg containing nitrides are the strongest candidates for FE polarization reversal (low barriers and high band gaps); yet, these compounds are NM.
In contrast, we found that MnSiN$_2$ and MnGeN$_2$ have robust G-AFM spin order but suboptimal semiconducting band gaps and, in the case of MnSiN$_2$, high reversal barriers, that will complicate ferroelectric switching. 
%
%

By strategically substituting alkali-earth metals like Zn and Mg at the Mn sites in MnSiN$_2$ and MnGeN$_2$, we designed ordered compounds that simultaneously exhibit strong FE and AFM behavior, reduced polarization reversal barriers, and enhanced insulating character. 
This dual optimization of electric and magnetic properties demonstrates a viable pathway for engineering altermagnetic, multiferroic nitrides with room-temperature stability. 
We recommend Zn/Mg-substituted ordered compounds of ternary manganese nitrides for future experimentation. 

Our design framework, grounded in electronegativity, bond ionicity, and defect-tolerant substitution, offers a generalizable strategy for discovering new quantum magnets. 
For example, applying these design principles to  rare-earth  (Eu,Mn)(Si/Ge)N$_2$ compounds, which host an $A$-site cation with an ionic radius comparable to Sr and incorporating localized $4f$ electrons, could yield rich and unconventional spin phenomena arising  from their interactions with Mn $3d$ electrons. 
These findings open promising avenues for the development of integrated AFM spintronic and multifunctional electronic devices.

\section{Computational Methods}
Density-functional-theory (DFT) computations were performed using the open-source Quantum Espresso \cite{Giannozzi2009, Giannozzi2017, Giannozzi2020} package with norm-conserving Vanderbilt pseudopotentials from the PseudoDojo library \cite{Setten2018, Lejaeghere2016, Hamann2013}. The generalized-gradient approximation (GGA) framework \cite{Perdew1992, Becke1988, Langreth1983} with the Perdew–Burke-Ernzerhof \cite{Perdew1996} exchange-correlation functionals was used. Spin-orbit coupling was not considered in this study. 
Although density functional theory (DFT) at the GGA level significantly underestimates the band gap \cite{Sham1983, Sham1985, Gorling2015}, our calculations aim to establish general trends within the $A$(Si/Ge)N$_2$ series.

The $k$-point spacing in the first Brillouin zone was set to 0.05 \AA$^{-1}$ and the kinetic energy cutoff was set to 90 Ry with a charge density cutoff of 360 Ry, so that the total energy and force were within 1.0 meV and 25 meV/\AA, respectively. The self-consistent field threshold was set to 10$^{-10}$ Ry. Geometry optimization was conducted for all crystal structures such that the total energy and force converged within 10$^{-5}$ Ry and 10$^{-4}$ Ry/bohr, respectively. The optimized structural parameters including the lattice constants and polar bond lengths are summarized in \supf1 of the SI. 
Band structures were computed based on the primitive orthorhombic lattice pathway \cite{Setyawan2010} while non-self-consistent-field calculations were conducted with a denser sampling of the first Brillouin zone at $k$-point spacing of 0.025 \AA$^{-1}$ to determine the PDOS (\supf2-6 of the SI). 

Four types of magnetic orders (one FM and three AFM) were considered to assess the stability of structures containing Mn. Although monoclinic MnSiN$_2$ with $Pc^\prime$ magnetic symmetry is demonstrated to exhibit an AFM canting angle of 0.6\textdegree\ \cite{Kautzsch2023}, we employ the orthorhombic MnSiN$_2$ and MnGeN$_2$ with $Pna2_1$ symmetry with the magnetization along the $\langle001\rangle$ crystallographic direction. Ordered compounds are generated based on the  MnSiN$_2$ and MnGeN$_2$ structures containing four formula units and substituting either one $A$ or $B$ site to model 25\% dopant molar concentration.

The dielectric function was calculated for MnSiN$_2$ and MnGeN$_2$ based on the independent-particle approximation \cite{Sipe1993}. The number of bands simulated were increased to a factor of two compared to the number of bands in the ground state, with a similar $k$-point sampling as the PDOS calculation (\supf7 and \supt2 of the SI). The minimum energy pathways of FE polarization reversal were calculated \textit{via} the nudged-elastic-band method, where a linear interpolation of images between the negatively-poled and positively-poled images is generated and the force orthogonal to the pathway is minimized \cite{Jonsson1998, Henkelman2000, Henkelman2000_2}. To generate the poled configurations, the $+P$ state was relaxed, and then the cations were translated along the polar $c$ axis, where the new structure is then relaxed and confirmed to be bistable. To capture the local switching behavior, a total of 51 images were generated while the self-consistent field threshold was reset to 10$^{-6}$ Ry to minimize the computational cost at reasonable accuracy. The electric polarization of each image was calculated using the Modern Theory of Polarization \textit{via} the Berry-phase method \cite{Resta2007, Spaldin2012, King-Smith1993}.  \\

\section*{Supporting Information}
\noindent Supporting Information is available from the Wiley Online Library or from the author.

\begin{acknowledgments}
\noindent This work was supported by the Air Force Office of Scientific Research under award number FA9550-23-1-0042.
The authors thank Ram Seshadri for useful discussions.
All first-principles calculations and analyses were conducted using the computational resources at the Quest high-performance computing facility at Northwestern University, which is jointly supported by the Office of the Provost, the Office for Research, and Northwestern University Information Technology.
\end{acknowledgments}

\section*{Conflict of Interest}
\noindent The authors declare no conflict of interest.

\section*{Author Contributions}
\noindent \textit{Conceptualization}--S. M. B., L.-D. Y., S. D. W., J. M. R.; \textit{methodology}--S. M. B., J. M. R.; 
\textit{software}--S. M. B., L.-D. Y., J. M. R.;
\textit{formal analysis}--S. M. B., L.-D. Y., J. M. R.; 
\textit{data curation}--S. M. B., L.-D. Y.; 
\textit{drafting}--S. M. B., L.-D. Y.; 
\textit{review and editing}--all authors; 
\textit{visualization}--S. M. B., L.-D. Y., J. M. R.; 
\textit{project administration}--J. M. R.; 
\textit{funding acquisition}--S. D. W., J. M. R.

\section*{Data Availability Statement}
\noindent The data that support the findings of this article are available in \cite{supp} and online at \cite{supp_dryad}.

\bibliography{references}

\end{document}


\title{{\sc supporting information}\\ \bf Ferroelectricity in antiferromagnetic wurtzite nitrides}

\author{Steven M.\ Baksa}
\affiliation{Department of Materials Science and Engineering, Northwestern University, Evanston, Illinois 60208, USA}

\author{Lin-Ding Yuan}
\affiliation{Department of Materials Science and Engineering, Northwestern University, Evanston, Illinois 60208, USA}

\author{Stephen D.\ Wilson}
\affiliation{Materials Department, University of California, Santa Barbara, Santa Barbara, California 93106, USA}

\author{James M.\ Rondinelli}
\email{jrondinelli@northwestern.edu}
\affiliation{Department of Materials Science and Engineering, Northwestern University, Evanston, Illinois 60208, USA}

\maketitle

\tableofcontents

\clearpage
\newpage

\section{Crystal structures}

\subsection{End-member structure analysis and validation}
\begin{figure}[h]
    \centering
    \includegraphics[scale=0.35]{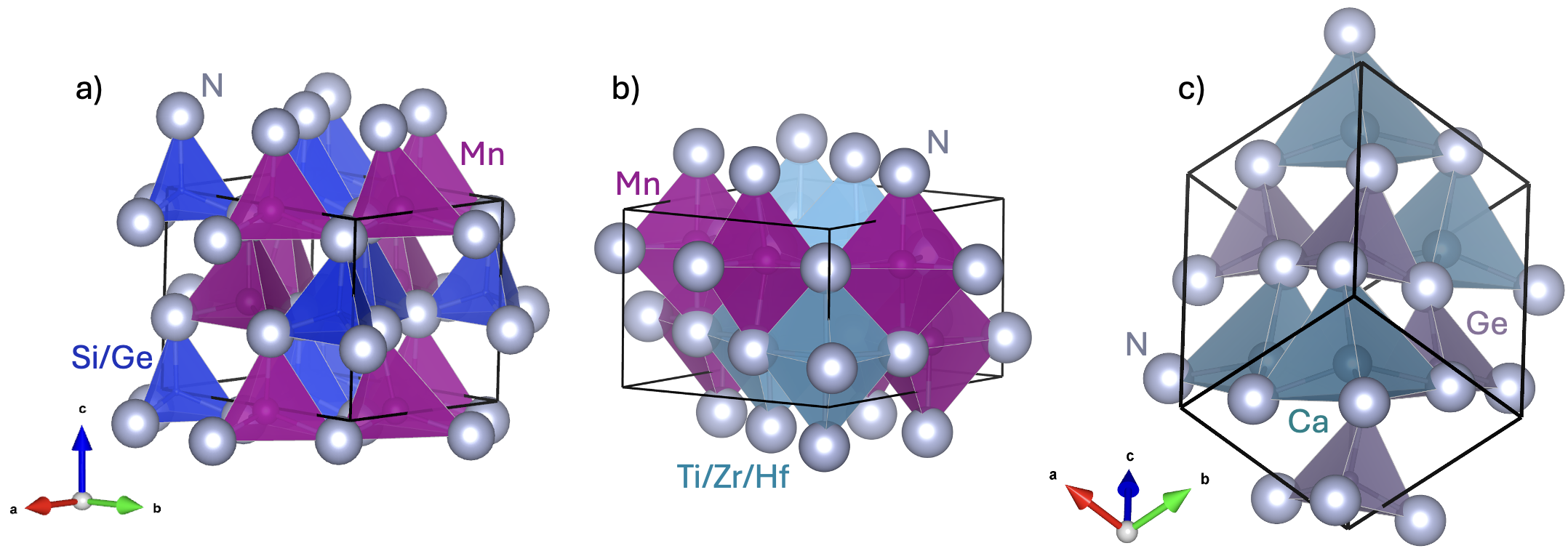}
    \caption{Crystal structures of (a) wurtzite-type, (b) centrosymmetric, and (c) zincblende-type end members. Here, Mn (dark purple), Si/Ge (dark blue), N (light gray), Ti/Zr/Hf (light blue), Ca (teal), and Ge (lilac) are indicated. Wurtzite-type and zincblende-type structures have net electric polarizations along the $\langle001\rangle$ and $\langle111\rangle$ crystallographic directions, respectively.}
    \label{fig:WZ-ZB}
\end{figure}

\autoref{fig:WZ-ZB} shows the wurtzite-type, centrosymmetric, and zincblende-type crystal structures of $AB$N$_2$. Here, the wurtzite-type structures are represented by Mn(Si/Ge)N$_2$ [\autoref{fig:WZ-ZB}(a)], the centrosymmetric structures are represented by Mn(Ti/Zr/Hf)N$_2$ [\autoref{fig:WZ-ZB}(b)],  and the singular zincblende-type structure is CaGeN$_2$ [\autoref{fig:WZ-ZB}(c)]. 
The wurtzite-type ($Pna2_1$) and centrosymmetric ($Pnma$) structures are orthorhombic, while the zincblende-type ($I\bar{4}2d$) structure is tetragonal. The $A$-site divalent ($e.g.,$ Mn$^{2+}$) and $B$-site tetravalent ($e.g.,$ Si$^{4+}$) cations are tetrahedrally coordinated by nitrogen anions (N$^{3-}$) in the wurtzite-type and zincblende-type structures, whereas these cations are five--coordinated (trigonal bipyramidal) by N$^{3-}$ in the centrosymmetric structures. 
The computed structural parameters of the end members, based on the DFT--PBE framework, are summarized in \autoref{tab:struc-param}, including the approximate $c/a$ ratio based on the conventional $c/a$ ratio in hexagonal wurtzites, where $\sim c/a = \frac{2c}{b}$. Although the computed structural parameters of nonmagnetic end members are within 1\% of experimental measurements, there exists a significant reduction in the unit-cell volume ($\Omega$) in MnSiN$_2$ and MnGeN$_2$ ($\sim$ 10\%) compared to the experimental values \cite{Bruls1999, Hausler2018, Kautzsch2023}. 
The centrosymmetric structures have no net electric polarization by definition. Wurtzite-type and zincblende-type structures have net electric polarizations along the $\langle001\rangle$ and $\langle111\rangle$ crystallographic directions, respectively. The effect of volume contraction on the ferroelectric (FE) reversal barrier was computed by scaling the images of the converged minimum-energy pathway to the experimental structural parameters and computing the resulting energies. The reversal barriers of MnSiN$_2$ and MnGeN$_2$ at the experimental structural parameters decreased by 11.8\% and increased by 1.2\%, respectively, compared to the reversal barriers from the PBE structural parameters. 

\begin{table}[!ht]
    \begin{center}
        \caption{Computed (DFT--PBE) structural parameters of the wurtzite-type, centrosymmetric, and zincblende-type end members [$AB$N$_2$ ($A$=Mn, Zn, Mg, Cd, Ca; $B$=Si, Ge, Ti, Zr, Hf)].}
        \label{tab:struc-param}
        \begin{ruledtabular}
        \begin{tabular}{llllllllll} 
             Material & Space group & $a$ (\AA) & $b$ (\AA) & $c$ (\AA) & $\sim c/a$ & $\Omega$ (\AA$^3$) & $r_{A\rm-N, polar}$ (\AA) & $r_{B\rm-N, polar}$ (\AA) \\ [0.5ex] 
             \hline
             MnSiN$_2$ & $Pna2_1$ & 5.191 & 6.251 & 4.790 & 1.533 & 155.42 & 1.97 & 1.75\\ 
             ZnSiN$_2$ & $Pna2_1$ & 5.285 & 6.308 & 5.068 & 1.607 & 168.94 & 2.07 & 1.76\\
             MgSiN$_2$ & $Pna2_1$ & 5.311 & 6.500 & 5.031 & 1.548 & 173.66 & 2.14 & 1.76\\
             CdSiN$_2$ & $Pna2_1$ & 5.432 & 6.753 & 5.299 & 1.569 & 194.38 & 2.29 & 1.76\\[0.3em]
             MnGeN$_2$ & $Pna2_1$ & 5.401 & 6.372 & 5.073 & 1.592 & 174.58 & 1.91 & 1.94\\ 
             ZnGeN$_2$ & $Pna2_1$ & 5.517 & 6.496 & 5.247 & 1.605 & 188.04 & 2.06 & 1.89\\
             MgGeN$_2$ & $Pna2_1$ & 5.555 & 6.667 & 5.229 & 1.569 & 193.63 & 2.12 & 1.89\\
             CdGeN$_2$ & $Pna2_1$ & 5.675 & 6.933 & 5.482 & 1.581 & 215.70 & 2.28 & 1.89\\[0.3em]
             MnTiN$_2$ & $Pnma$ & 5.701 & 6.896 & 4.061 & 1.178 & 159.66 & 2.05 & 2.03\\
             MnZrN$_2$ & $Pnma$ & 5.997 & 7.082 & 4.340 & 1.226 & 184.33 & 2.20 & 2.17\\
             MnHfN$_2$ & $Pnma$ & 5.975 & 7.051 & 4.292 & 1.217 & 180.80 & 2.18 & 2.15\\[0.3em]
             CaGeN$_2$ & $I\bar{4}2d$ & 5.310 & 6.558 & 5.309 & 1.619 & 184.89 & 2.42 & 1.89\\
        \end{tabular}
        \end{ruledtabular}
    \end{center}
\end{table}
\vspace{-12pt}
\enlargethispage{2em}

\clearpage
\subsection{Cation-ordered nitrides}
The computed structural, FE, and magnetic properties of the ordered nitride compounds, based on the DFT--PBE framework, are outlined in \autoref{tab:ordered-struc-param}, which include the FE reversal barrier $\Delta$, the magnetic ground state (GS), and the absolute value of the average local magnetic moment centered on the Mn sites $|m|$. In the case of the $A$-site ordered nitrides, the $\Delta$ values range from 0.469 eV/f.u. for Mn$_3$MgGe$_4$N$_8$ to 0.611 eV/f.u. for Mn$_3$CdSi$_4$N$_8$. In the case of the $B$-site ordered nitrides, the $\Delta$ values range from 0.474 eV/f.u. for Mn$_4$Ge$_3$ZrN$_8$ to 0.701 eV/f.u. for Mn$_4$Ge$_4$HfN$_8$. The magnetic GS of the A-site and B-site ordered compounds are ferrimagnetic (FiM) and G-type antiferromagnetic (G-AFM), respectively. The $|m|$ of the $A$-site and $B$-site ordered nitrides are centered at 4.60 and 4.36 $\mu_{\rm B}$, respectively (see Sect. IV for discussion of the magnetic properties).

\begin{table}[!ht]
    \begin{center}
        \caption{Computed (DFT-PBE) structural, ferroelectric, and magnetic properties of the ordered compounds. All ordered compounds have $P1$ symmetry.  $|m|$ refers to the absolute value of the average local magnetic moment.}
        \label{tab:ordered-struc-param}
        \begin{ruledtabular}
        \begin{tabular}{llllllllll} 
             Material & $a$ (\AA) & $b$ (\AA) & $c$ (\AA) & $\sim c/a$ & $\Omega$ (\AA$^3$) & $\Delta$ (eV/f.u.) & Magnetic GS & $|m|$ ($\mu_{\rm B}$) \\ [0.5ex] 
             \hline
             Mn$_3$ZnSi$_4$N$_8$ & 5.265 & 6.414 & 5.096 & 1.589 & 172.07 & 0.599 & FiM & 4.56\\ 
             Mn$_3$MgSi$_4$N$_8$ & 5.271 & 6.465 & 5.082 & 1.572 & 173.19 & 0.528 & FiM & 4.56\\
             Mn$_3$CdSi$_4$N$_8$ & 5.298 & 6.529 & 5.149 & 1.577 & 178.09 & 0.611 & FiM & 4.60\\[0.3em]
             Mn$_3$ZnGe$_4$N$_8$ & 5.499 & 6.594 & 5.286 & 1.603 & 191.69 & 0.505 & FiM & 4.59\\
             Mn$_3$MgGe$_4$N$_8$ & 5.507 & 6.641 & 5.279 & 1.590 & 193.07 & 0.469 & FiM & 4.60 \\ 
             Mn$_3$CdGe$_4$N$_8$ & 5.535 & 6.712 & 5.339 & 1.591 & 198.34 & 0.512 & FiM & 4.62 \\
             Mn$_3$CaGe$_4$N$_8$ & 5.546 & 6.831 & 5.304 & 1.553 & 200.98 & 0.476 & FiM & 4.64 \\[0.3em]
             Mn$_4$Si$_3$TiN$_8$ & 5.343 & 6.444 & 5.151 & 1.599 & 177.36 & 0.582 & G-AFM & 4.35 \\
             Mn$_4$Si$_3$ZrN$_8$ & 5.416 & 6.512 & 5.171 & 1.588 & 182.35 & 0.574 & G-AFM & 4.34 \\
             Mn$_4$Si$_3$HfN$_8$ & 5.408 & 6.504 & 5.175 & 1.591 & 182.04 & 0.576 & G-AFM & 4.35 \\[0.3em]
             Mn$_4$Ge$_3$TiN$_8$ & 5.517 & 6.580 & 5.306 & 1.613 & 192.65 & 0.687 & G-AFM & 4.38 \\
             Mn$_4$Ge$_3$ZrN$_8$ & 5.597 & 6.640 & 5.334 & 1.607 & 198.23 & 0.474 & G-AFM & 4.37 \\
             Mn$_4$Ge$_3$HfN$_8$ & 5.588 & 6.635 & 5.336 & 1.609 & 197.84 & 0.701 & G-AFM & 4.39 \\
        \end{tabular}
        \end{ruledtabular}
    \end{center}
\end{table}

\clearpage

\section{Electronic structures}
\begin{figure}[ht]
    \centering
    \includegraphics[width=0.9\linewidth]{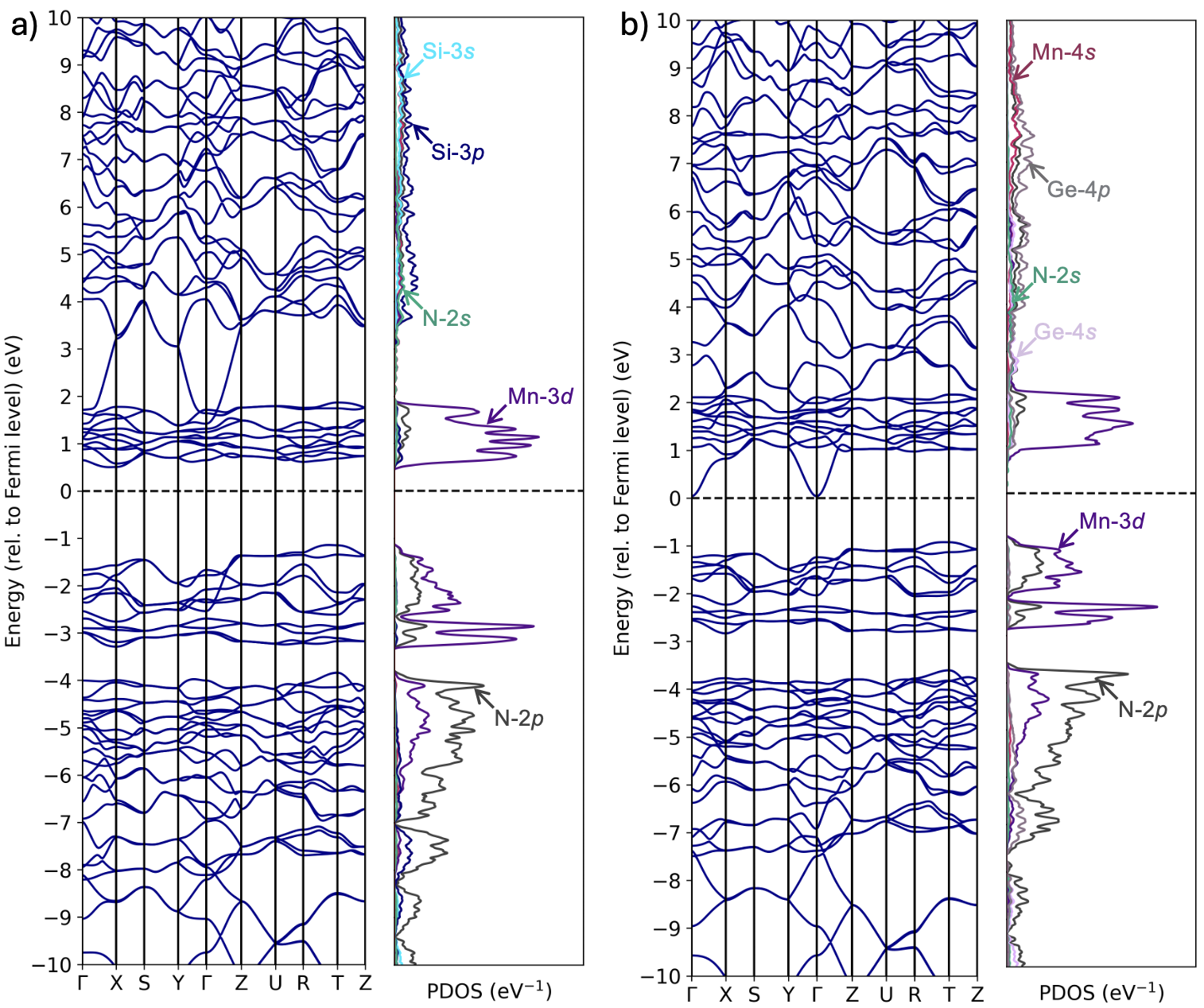}
    \caption{The band structure and projected density of states (PDOS) of G-AFM (a) MnSiN$_2$ and (b) MnGeN$_2$ computed based on the PBE exchange-correlation functional \cite{Perdew1996}. The N-2$s$ (light teal), N-2$p$ (dark gray), Mn-3$d$ (purple), Mn-4$s$ (ruby), Si-3$s$ (sky blue), Si-3$p$ (dark blue), Ge-4$s$ (lavender), and Ge-4$p$ (dark lilac) orbitals are indicated. All energies are relative to the Fermi energy (dotted line). The band gap drops from 1.64 to 0.96 eV as the $B$ site changes from Si to Ge, respectively, with the conduction band edge in MnGeN$_2$ just above the Fermi level. The band edges are largely dominated by the Mn-3$d$ and N-2$p$ orbitals.}
    \label{fig:PDOS-MnSiN2-MnGeN2}
\end{figure}

\begin{figure}
    \centering
    \includegraphics[width=0.9\linewidth]{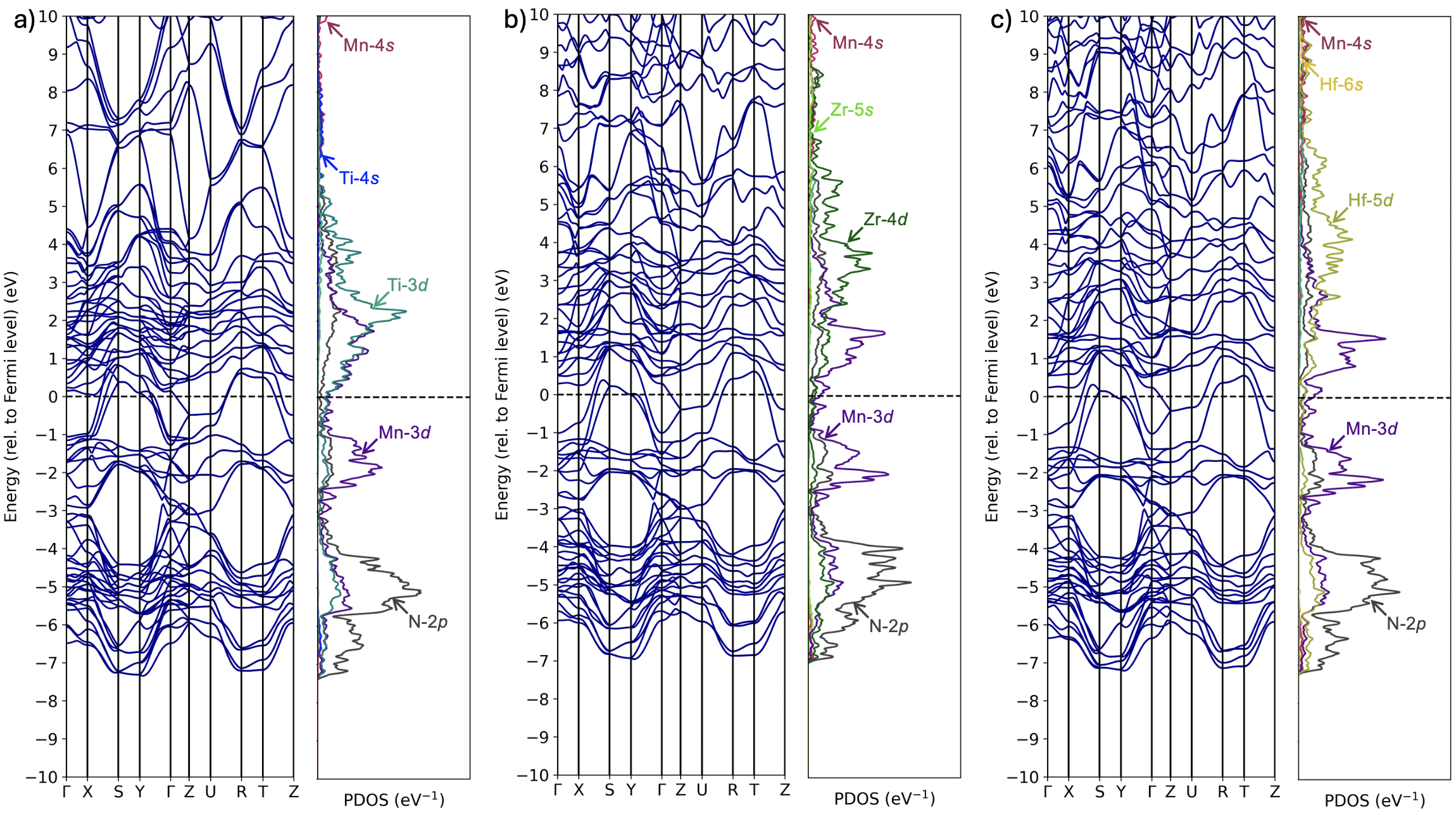}
    \caption{The band structure and projected density of states (PDOS) of centrosymmetric (a) MnTiN$_2$, (b) MnZrN$_2$, and (c) MnHfN$_2$ computed based on the PBE exchange-correlation functional \cite{Perdew1996}. The N-2$s$ (light teal), N-2$p$ (dark gray), Mn-3$d$ (purple), Mn-4$s$ (ruby), Ti-3$d$ (teal), Ti-4$s$ (dark blue), Zr-4$d$ (dark green), Zr-5$s$ (light green), and Hf-5$d$ (chartreuse), and Hf-6$s$ (gold) orbitals are indicated. All energies are relative to the Fermi energy (dotted line). Under the DFT-PBE framework, these centrosymmetric end members appear metallic with no band gap.}
    \label{fig:PDOS-centrosymmetric}
\end{figure}

\begin{figure}
    \centering
    \includegraphics[width=0.9\linewidth]{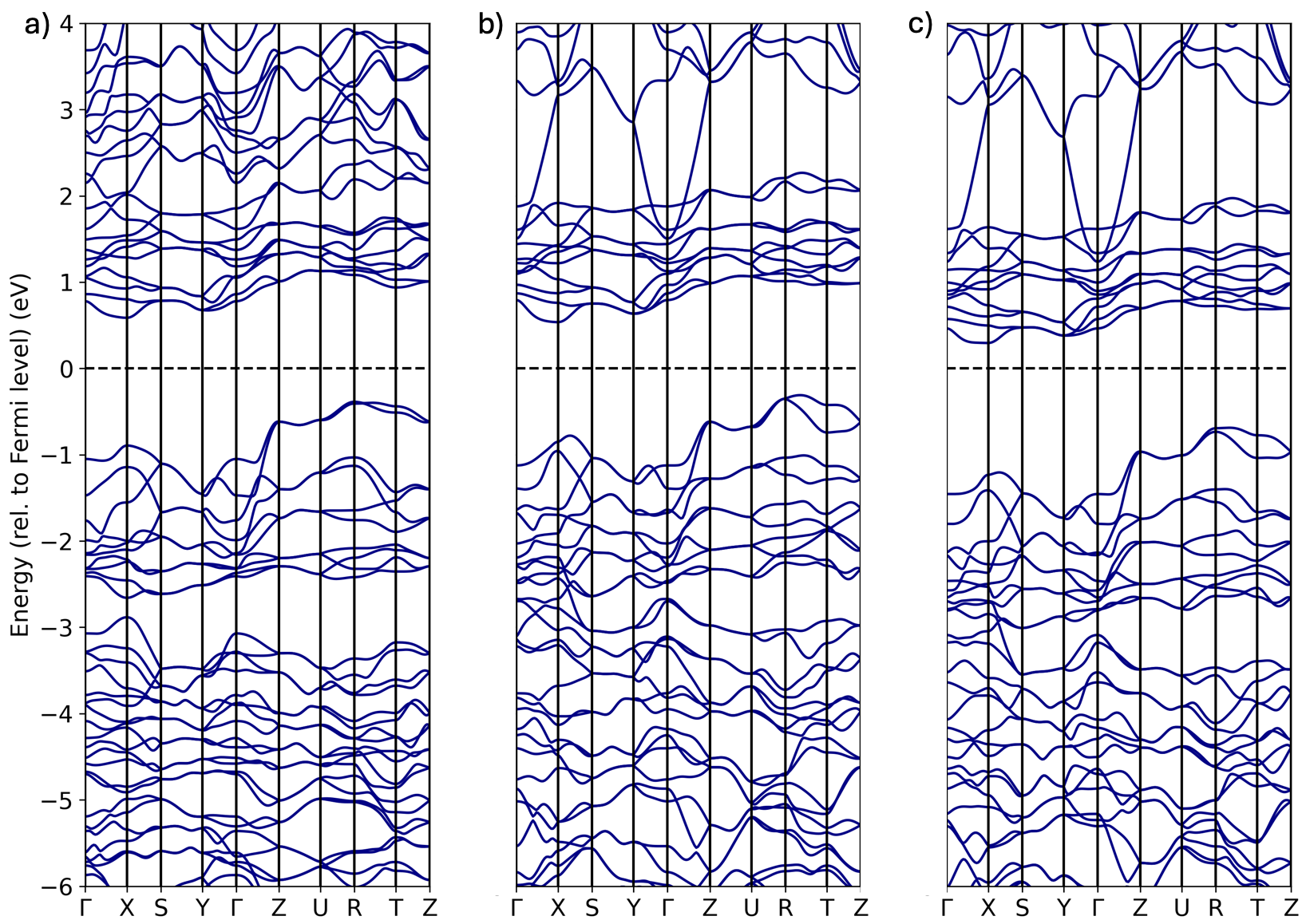}
    \caption{The band structure and projected density of states (PDOS) of polar, wurtzite-type (a) MnTiN$_2$, (b) MnZrN$_2$, and (c) MnHfN$_2$ computed based on the PBE exchange-correlation functional \cite{Perdew1996}. All energies are relative to the Fermi energy (dotted line). Significant band gaps exist in the polar phases compared to the metallic behavior of the centrosymmetric phases under the DFT-PBE framework. The simulated band structures exhibit features similar to those computed for MnSiN$_2$ and MnGeN$_2$.}
    \label{fig:band-structures-centro-polar}
\end{figure}

\begin{figure}
    \centering
    \includegraphics[width=0.9\linewidth]{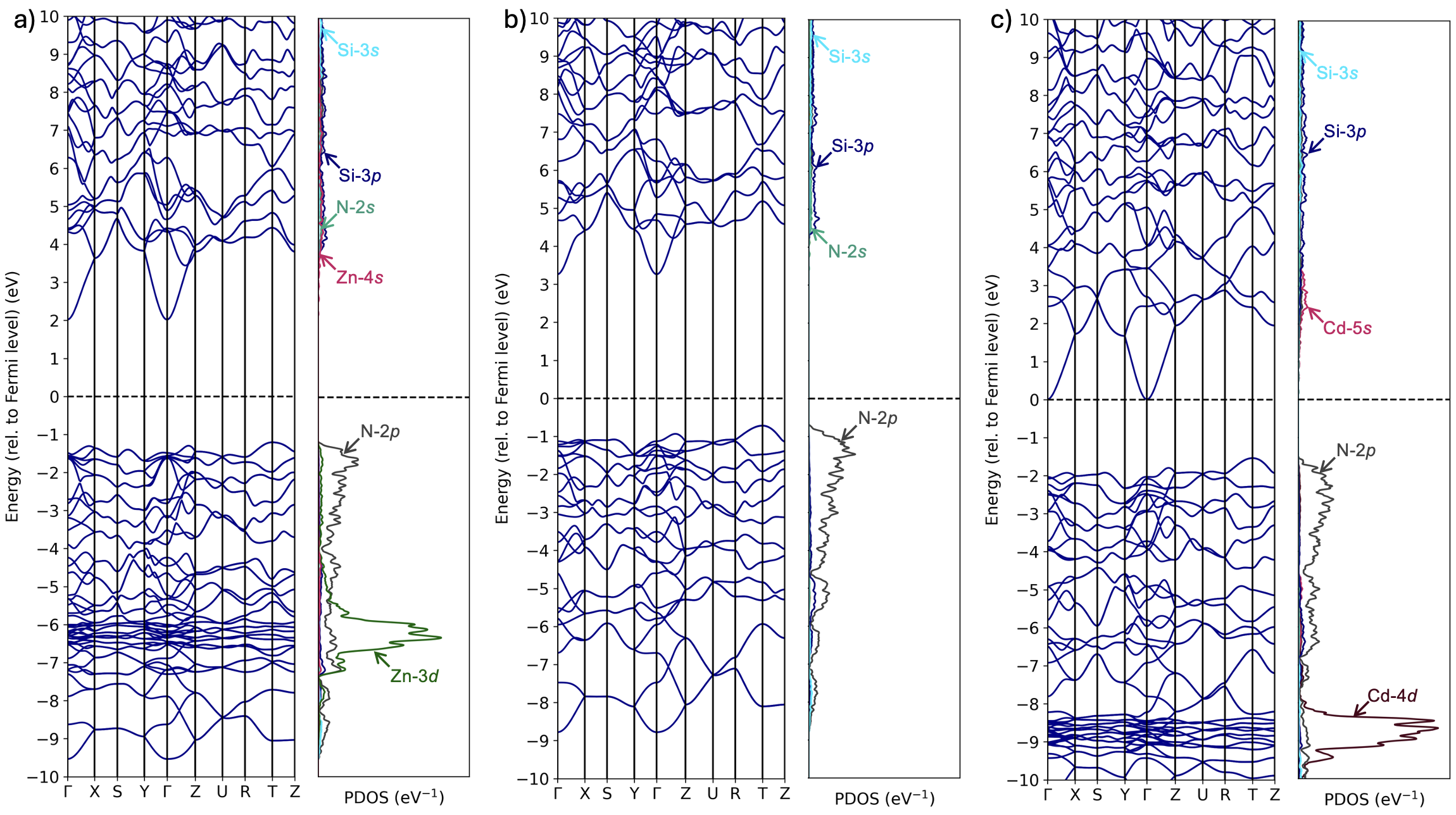}
    \caption{The band structure and projected density of states (PDOS) of nonmagnetic (a) ZnSiN$_2$, (b) MgSiN$_2$, and (c) CdSiN$_2$ computed based on the PBE exchange-correlation functional \cite{Perdew1996}. The N-2$s$ (light teal), N-2$p$ (dark gray), Si-3$s$ (sky blue), Si-3$p$ (dark blue), Zn-3$d$ (dark green), Zn-4$s$ (magenta), Cd-4$d$ (dark brown), and Cd-5$s$ (magenta) orbitals are indicated. The Mg orbitals are not visible due to low intensity. All energies are relative to the Fermi energy (dotted line). MgSiN$_2$ had the largest band gap at 3.98 eV, while CdSiN$_2$ had the lowest band gap at 1.54 eV with the conduction band edge just above the Fermi level. The band edges are largely dominated by the $A$-site-$s$ and N-2$p$ orbitals.}
    \label{fig:PDOS-nonmagnetic-SiN}
\end{figure}

\begin{figure}
    \centering
    \includegraphics[width=0.9\linewidth]{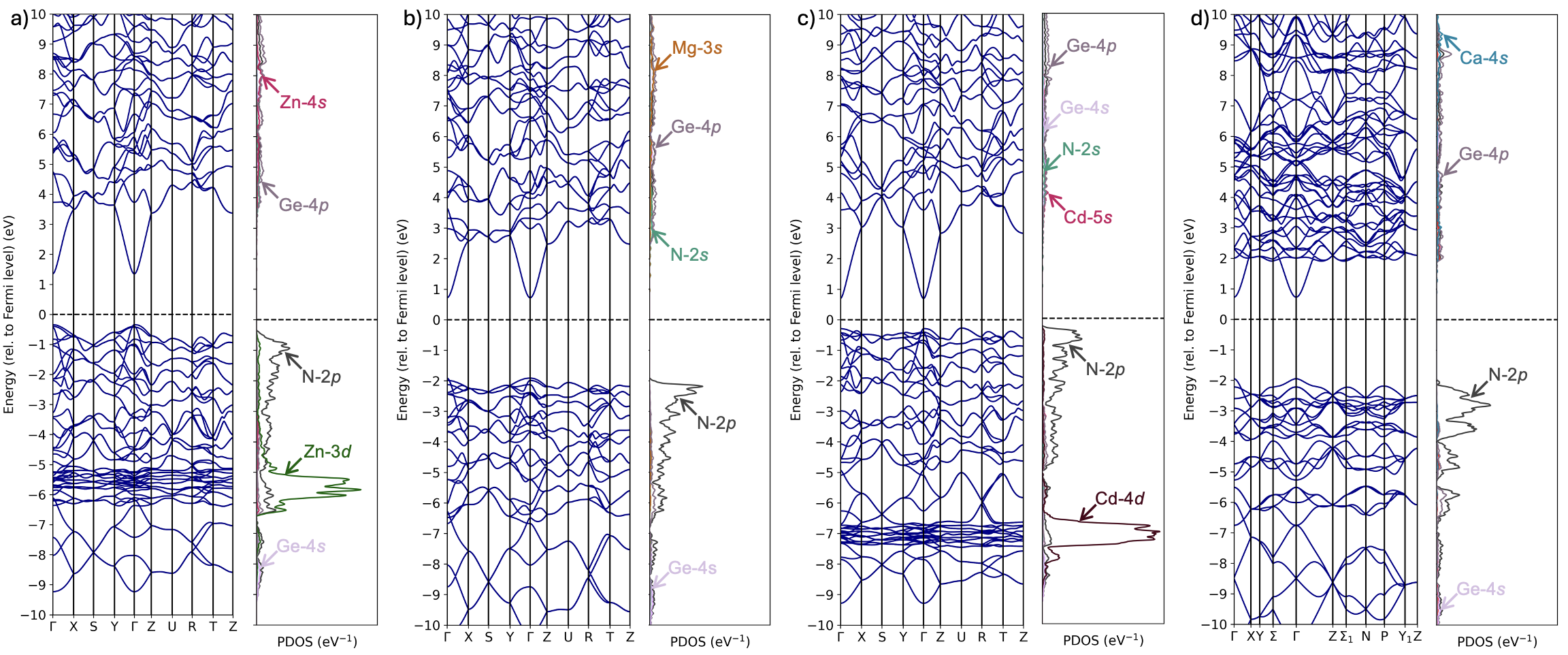}
    \caption{The band structure and projected density of states (PDOS) of nonmagnetic (a) ZnGeN$_2$, (b) MgGeN$_2$, (c) CdGeN$_2$, and (d) CaGeN$_2$ computed based on the PBE exchange-correlation functional \cite{Perdew1996}. While ZnGeN$_2$, MgGeN$_2$, and CdGeN$_2$ have a wurtzite-type structure, CaGeN$_2$ has a zincblende-type structure. The N-2$s$ (light teal), N-2$p$ (dark gray), Ge-4$s$ (lavender), Ge-4$p$ (dark lilac), Zn-3$d$ (dark green), Zn-4$s$ (magenta), Mg-3$s$ (orange), Cd-4$d$ (dark brown), Cd-5$s$ (magenta), and Ca-4$s$ (teal) orbitals are indicated. All energies are relative to the Fermi energy (dotted line). CaGeN$_2$ had the largest band gap at 2.67 eV while CdGeN$_2$ had the lowest band gap at 0.97 eV. The band edges are largely dominated by the N-2$p$ orbitals.}
    \label{fig:PDOS-nonmagnetic-GeN}
\end{figure}

The band structures and projected density of states (PDOS) have been computed for all end members ($AB$N$_2$) (Figs. \ref{fig:PDOS-MnSiN2-MnGeN2}-\ref{fig:PDOS-nonmagnetic-GeN}) \textit{via} the DFT-PBE framework \cite{Perdew1996}. In the case of G-type antiferromagnetic (G-AFM) wurtzite-type MnSiN$_2$ and MnGeN$_2$ (cf. \autoref{fig:PDOS-MnSiN2-MnGeN2}a,b), the N-2$s$ (light teal), N-2$p$ (dark gray), Mn-3$d$ (purple), Mn-4$s$ (ruby), Si-3$s$ (sky blue), Si-3$p$ (dark blue), Ge-4$s$ (lavender), and Ge-4$p$ (dark lilac) orbitals are indicated. The band gap drops from 1.64 to 0.96 eV as the $B$ site changes from Si to Ge, respectively. The band edges are largely dominated by the Mn-3$d$ and N-2$p$ orbitals. 

In the case of G-AFM centrosymmetric MnTiN$_2$, MnZrN$_2$, and MnHfN$_2$, the Ti-3$d$ (teal), Ti-4$s$ (dark blue), Zr-4$d$ (dark green), Zr-5$s$ (light green), and Hf-5$d$ (chartreuse), and Hf-6$s$ (gold) orbitals are indicated. The bands near the Fermi level are largely dominated by the Mn-3$d$ and $B$-site-$d$ orbitals. Although the band structures at the PBE level exhibit metallic behavior with no band gap, there is the possibility of semiconducting behavior in these centrosymmetric end members, given the systematic underestimation of the band gap in the PBE framework \cite{Sham1983, Sham1985, Gorling2015} and the low PDOS intensities near the Fermi energy. In addition, two cases were considered to further understand the metallic states for MnTiN$_2$, MnZrN$_2$, and MnHfN$_2$:  (1) An onsite Hubbard $U$ parameter is applied to the Mn-3$d$ orbitals ($U = 5$ eV) to account for self-interaction errors, and (2) Inversion symmery breaking is included by computing MnTiN$_2$, MnZrN$_2$, and MnHfN$_2$ in the polar, wurtzite-type phase. 

With the application of the Hubbard $U$ parameter to the centrosymmetric phases of MnTiN$_2$, MnZrN$_2$, and MnHfN$_2$, metallic character persisted, indicating a potential structural mechanism. 
The band structures of the polar phases of MnTiN$_2$, MnZrN$_2$, and MnHfN$_2$ illustrate their semiconducting behavior compared to the centrosymmetric phases (\autoref{fig:band-structures-centro-polar}). This substantive change in optical behavior is consistent with the chemical bonding of these two phases \cite{Kawamura2022}. Recall that the polar phases contain Mn$^{2+}$ and B$^{4+}$ cations that are tetrahedrally coordinated by N$^{3-}$ anions, while the cations in the centrosymmetric phases are in a penta-coordinated trigonal bipyramidal environment. The bandwidth and overlap of the filled electron orbitals generally increase with the cation coordination number, diminishing the overall band gap. Bond lengths also affect the band gap, where short bonds are generally strong bonds, leading to large band gaps. The Mn-N bond lengths decrease approximately from 2.2~\AA\ in the centrosymmetric phase to 1.9~\AA\ in the polar phase, consistent with the metallic-semiconductor transition. We note that the band structures of the polar phases of MnTiN$_2$, MnZrN$_2$, and MnHfN$_2$ exhibit features similar to those observed in MnSiN$_2$ and MnGeN$_2$, including the elevated valence band on the Z-U-R-T section of the reciprocal-space pathway and the pseudogap at approximately 3 eV below the Fermi energy. 

In the case of the non-magnetic wurtzite-type ZnSiN$_2$, MgSiN$_2$, and CdSiN$_2$ (cf. \autoref{fig:PDOS-nonmagnetic-SiN}a--c), the Zn-3$d$ (dark green), Zn-4$s$ (magenta), Cd-4$d$ (dark brown), and Cd-5$s$ (magenta) orbitals are indicated. MgSiN$_2$ had the largest band gap at 3.98 eV, while CdSiN$_2$ had the lowest band gap at 1.54 eV. It is noted that the conduction band edge of CdSiN$_2$ lies just a few tenths of an eV above the Fermi level, hinting at a possible transition to a metallic state. The band edges are largely dominated by the $A$-site-$s$ and N-2$p$ orbitals. Among the non-magnetic wurtzite-type ZnGeN$_2$, MgGeN$_2$, and CdGeN$_2$ (cf. \autoref{fig:PDOS-nonmagnetic-GeN}a--c), the MgGeN$_2$ had the largest band gap at 2.63 eV while CdGeN$_2$ had the smallest band gap at 0.97 eV. In contrast, the non-magnetic tetragonal zincblende-type CaGeN$_2$ (\autoref{fig:PDOS-nonmagnetic-GeN}d) had the largest band gap at 2.67 eV, even among the $A$GeN$_2$ series. The valence band edges are largely dominated by the N-2$p$ orbitals with low intensity from the conduction band edges. Although the Mg orbitals are not visible in the PDOS of MgSiN$_2$ due to low intensity, the Mg-3$s$ orbitals have intriguingly significant intensities in the PDOS of MgGeN$_2$.

\newpage
\clearpage

\section{Dielectric properties}
\begin{figure}[ht]
    \centering
    \includegraphics[scale=0.50]{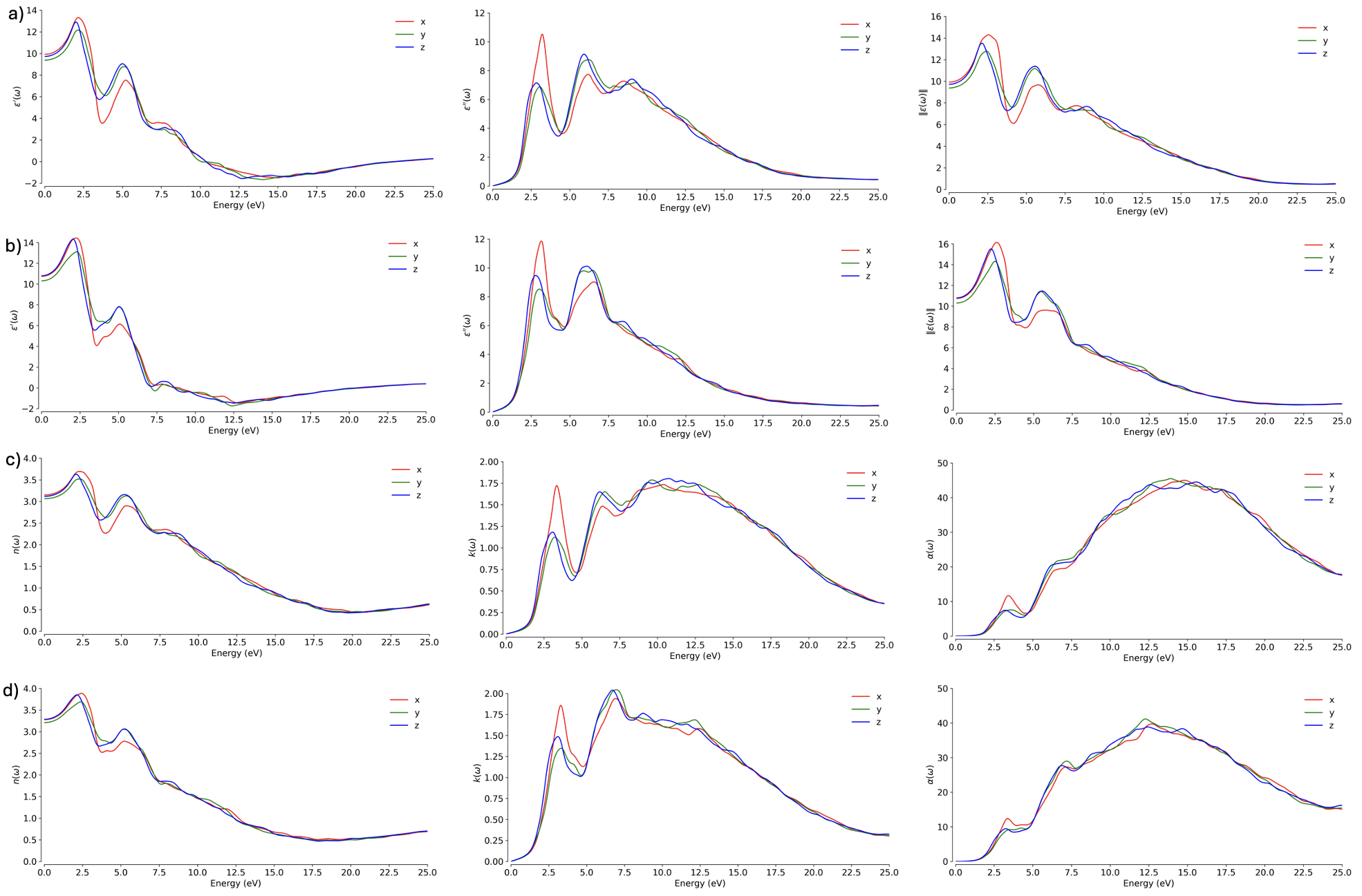}
    \caption{Dielectric function of G-AFM (a) MnSiN$_2$ and (b) MnGeN$_2$. The real component [$\epsilon'(\omega)$, left], imaginary component [$\epsilon''(\omega)$, middle], and magnitude [$||\epsilon (\omega)||$, right] of the dielectric function are computed based on the PBE exchange-correlation functional \cite{Perdew1996} and the independent-particle approximation \cite{Sipe1993}. The derived dielectric properties of (c) MnSiN$_2$ and (d) MnGeN$_2$. These properties include the refractive index [$n(\omega)$, left], the extinction coefficient [$k(\omega)$, middle], and the linear absorption coefficient [$\alpha(\omega)$, right]. The dielectric function and properties are projected along the Cartesian directions with $x$ as red, $y$ as green, and $z$ as blue.}
    \label{fig:Dielectric-panel}
\end{figure}

The dielectric functions of G-AFM MnSiN$_2$ and MnGeN$_2$ are computed based on the DFT-PBE framework \cite{Perdew1996} and the independent-particle approximation \cite{Sipe1993} are shown in \autoref{fig:Dielectric-panel}a,b. The dielectric function is 
\begin{equation}
    \epsilon_i(\omega) = \epsilon'_i(\omega)+i\epsilon''_i(\omega)
    \label{eqn:dielectric-function}
\end{equation}
where 
$\epsilon'$ is the real component, $\epsilon''$ is the imaginary component, and $\omega$ is the frequency of the applied electric field. Since $\epsilon$ is a tensor, the components $x,y,z$ are denoted by $i$. The magnitude of the dielectric function is calculated based on the real and imaginary components as: 
\begin{equation}
    ||\epsilon_i(\omega)|| = \sqrt{\epsilon'_i(\omega)^2+\epsilon''_i(\omega)^2}\,.
    \label{eqn:dielectric-magnitude}
\end{equation}
The dielectric constant ($i.e.,$ the dielectric function near zero frequency) of MnSiN$_2$ ($\epsilon=9.67$) is lower than that of MnGeN$_2$ ($\epsilon=10.61$), which is consistent with MnSiN$_2$ having a larger band gap than MnGeN$_2$. The dielectric function is also projected along the Cartesian directions with $x$ as red, $y$ as green, and $z$ as blue. Anisotropy of the dielectric function is observed as expected from an orthorhombic crystal system.

\renewcommand{\arraystretch}{1.2}
\begin{table}[ht]
    \begin{center}
        \caption{Computed and experimental optical and dielectric properties of the semiconducting ternary nitride end members. These properties include the band gap, dielectric constant, and the electro-optic coefficients.}
        \label{tab:dielectric}
        \begin{ruledtabular}
        \begin{tabular}{l l l l l l l l  l } 
             Material & $E_{\rm g, DFT}$ (eV) & $E_{\rm g, Expt.}$ (eV) & $\epsilon_{\rm 0,DFT}$ & $\epsilon_{\rm 0,Expt.}$ & $n_0$ & $\Delta n_0$ & $r_{13}$ (pm/V) & $r_{33}$ (pm/V) \\ [0.5ex] 
             \hline
             MnSiN$_2$ & 1.64 & 2.3-3.5 \cite{Hausler2018, Kautzsch2023} & 9.67 & - & 3.11 & 0.54 & -& - \\ 
             ZnSiN$_2$ & 3.23 & 3.7-4.2 \cite{Hausler2017, Ogura2021} & 6.78 & 5.0 \cite{Mintairov2000} & 2.60 & 0.25 & 5.77 &-16.89\\
             MgSiN$_2$ & 3.98 & 4.8 \cite{Hausler2018, Fang1999} & 5.39 & 10.5-15 \cite{Groen1993, Kageyama2025} & 2.32 & 0.13 & -2.26 &-2.45 \\
             CdSiN$_2$ & 1.54 & - & 7.23 & - & 2.69 & 0.54 & 2.74 & 35.62\\[0.3em]
             MnGeN$_2$ & 0.96 & 2.5 \cite{Hausler2018} & 10.61 & - & 3.26 & 0.50 & - & -\\
             ZnGeN$_2$ & 1.69 & 2.7-3.7 \cite{Hausler2017, Ogura2021} & 8.08 & 5.2-6.3 \cite{Misaki2003} & 2.84 & 0.16 & -17.52 & 36.68\\
             MgGeN$_2$ & 2.63 & 3.2 \cite{Hausler2018} & 6.23 & - & 2.50 & 0.18 & -11.70 & 17.91\\
             CdGeN$_2$ & 0.97 & - & 9.11 & - & 3.02 & 0.03 & -26.73 & 33.16\\[0.3em]
             CaGeN$_2$ & 2.67 & - & 6.93 & - & 2.63 & 0.25 & 0 & 0\\
        \end{tabular}
        \end{ruledtabular}
    \end{center}
\end{table}
Critical dielectric properties can be derived from the dielectric function \cite{Bhuyan2017} and Equations \ref{eqn:refractive-index}, \ref{eqn:extinction-coefficient}, and \ref{eqn:absorption-coefficient}:
\begin{equation}
    n_i(\omega) = \sqrt{\frac{||\epsilon_i(\omega)||+\epsilon'_i(\omega)}{2}}
    \label{eqn:refractive-index}
\end{equation}
\begin{equation}
    k_i(\omega) = \sqrt{\frac{||\epsilon_i(\omega)||-\epsilon'_i(\omega)}{2}}
    \label{eqn:extinction-coefficient}
\end{equation}
\begin{equation}
    \alpha_i(\omega) = \sqrt{2} \omega \sqrt{||\epsilon_i(\omega)||-\epsilon_i'(\omega)}=2\omega k_i(\omega)
    \label{eqn:absorption-coefficient}
\end{equation}
where $n$ is the refractive index, $k$ is the extinction coefficient, and $\alpha$ is the linear absorption coefficient. These properties are computed for MnSiN$_2$ and MnGeN$_2$ (cf.\ \autoref{fig:Dielectric-panel}c,d). 

Consistent with the dielectric constant, the refractive index of MnSiN$_2$ ($n \sim 3.1$) is smaller than that of MnGeN$_2$ ($n \sim 3.3$). \autoref{tab:dielectric} summarizes the optical and dielectric properties of the semiconducting end members. As expected, the GGA-PBE framework significantly underestimates (overestimates) the band gap (dielectric constant) in comparison to experiment. One noticeable exception is MgSiN$_2$, where the computed band gap and dielectric constant are underestimated. This discrepancy can be rationalized due to significant discrepancies in the measured dielectric constant \cite{Groen1993, Kageyama2025}. 

The electro-optic coefficients are measures of the response of $n$ with respect to an incoming electric field: 
\begin{equation}
    \Delta \left(\frac{1}{n^2}\right)_{ij} = \sum_{k}^{xyz} r_{ij}E_k
    \label{eqn:EO}
\end{equation}
where $r_{ij}$ is the electro-optic coefficient matrix and $E_k$ is the applied electric field. Magnetic members were not considered since a spin-polarized case of the calculation of the electro-optic coefficient has not been implemented in Quantum Espresso \cite{Giannozzi2009, Giannozzi2017, Giannozzi2020}. Current implementations of Quantum Espresso compute the electro-optic coefficient based on the local-density-approximation (LDA) framework \cite{Hohenberg1964}. Except for zincblende-type CaGeN$_2$, which showed zero electro-optic coupling, computed values of $r_{ij}$ are on the order of 2-37 pm/V. The magnitudes of the $r_{13}$ coefficients ranged from 2.3 pm/V for MgSiN$_2$ to 26.7 pm/V for CdGeN$_2$, whereas the magnitudes of the $r_{33}$ coefficients ranged from 2.5 pm/V for MgSiN$_2$ to 36.7 pm/V for ZnGeN$_2$. In comparison, these computed values are generally an order of magnitude larger than computed electro-optic coefficient values of substituted nitrides like (Al,Sc)N ($\sim$5 pm/V) \cite{Pockels-Yang2024}. These computed electro-optic coefficients, however, are similar in magnitude to electro-optic measurements observed in LiNbO$_3$ and PbTiO$_3$ and an order of magnitude below electro-optic measurements in state-of-the-art BaTiO$_3$ \cite{Wen2024}.

\newpage

\section{Magnetic properties}
The magnetic properties of the end members and ordered compounds are considered based on the Heisenberg spin model: 
\begin{equation}
    E = E_0-J \sum_{i>j} \vec{S}_i \cdot \vec{S}_j
    \label{eqn:Heisenberg}
\end{equation}
Here, $J$ is the magnetic exchange parameter where $J>0$ is ferromagnetic and $J<0$ is antiferromagnetic, $\vec{S}_i$ is the spin of site $i$, $E_0$ is the energy of the material system with no spin interactions, and $E$ is the energy per magnetic site. The main text provides the unit cell of G-AFM MnSiN$_2$ with the $J_{\rm v}$ and $J_\parallel$ magnetic interactions labeled. 

Equations \ref{eqn:FM}-\ref{eqn:G-FM} provide a 
system of equations to solve for the exchange parameters with the assumption that $S=5/2$ and the spins are along $\langle001\rangle$:
\begin{equation}
    E_{\rm FM} = E_0 - 2J_{\rm v}S^2 - 2J_{\parallel}S^2
    \label{eqn:FM}
\end{equation}
\begin{equation}
    E_{\rm A-AFM} = E_0 + 2J_{\rm v}S^2 - 2J_{\parallel}S^2
    \label{eqn:A-FM}
\end{equation}
\begin{equation}
    E_{\rm C-AFM} = E_0 - 2J_{\rm v}S^2 + 2J_{\parallel}S^2
    \label{eqn:C-FM}
\end{equation}
\begin{equation}
    E_{\rm G-AFM} = E_0 + 2J_{\rm v}S^2 + 2J_{\parallel}S^2
    \label{eqn:G-FM}
\end{equation}

\begin{figure}[ht]
    \centering
    \includegraphics[scale=0.40]{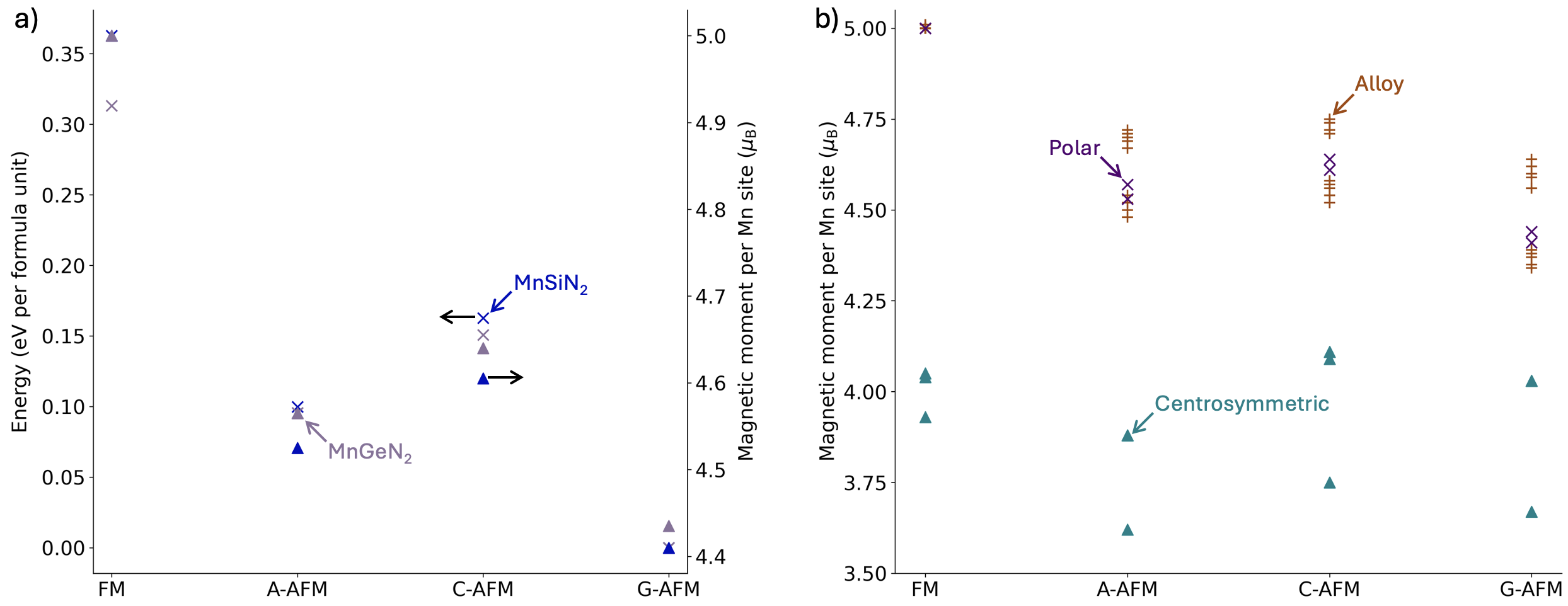}
    \caption{(a) Benchmarking of magnetic orders of polar MnSiN$_2$ (blue) and MnGeN$_2$ (lilac). The ferromagnetic (FM) and three antiferromagnetic (AFM) cases were considered. In both end members, the G-AFM case was the most stable with the lowest magnetic moment at approximately 4.4 $\mu_{\rm B}$ per Mn site, whereas the FM case was the least stable with the highest magnetic moment at 5.0 $\mu_{\rm B}$ per Mn site. (b) Magnetic moments of polar (purple cross), centrosymmetric (teal triangle), and ordered compounds at 25\% dopant molar concentration (orange plus) as a function of the magnetic ordering. The polar and ordered compounds tend to have larger magnetic moments than centrosymmetric structures, regardless of magnetic ordering. The magnetic moments of ordered compounds are centered around that of the polar end member with little deviation. On average, the G-AFM magnetic ordering had the lowest magnetic moment for the polar and ordered compounds, while the A-AFM magnetic ordering had the lowest magnetic moment for the centrosymmetric end members.}
    \label{fig:Magnetic-benchmarking}
\end{figure}

\noindent The magnetic orders in MnSiN$_2$ and MnGeN$_2$ are benchmarked based on their energetics and magnetic moment per Mn site (cf. \autoref{fig:Magnetic-benchmarking}a). A total of four magnetic orderings were considered: (1) ferromagnetic (FM), (2) A-type (A-AFM), (3) C-type (C-AFM), and (4) G-type (G-AFM). For both MnSiN$_2$ (blue) and MnGeN$_2$ (lilac), the G-AFM ordering was the most stable with the lowest magnetic ordering ($\sim$ 4.4 $\mu_{\rm B}$ per Mn site). The order of stability for the magnetic configurations from most to least stable is G-AFM, A-AFM, C-AFM, and FM. Note that  FM ordering is at least 0.30 eV per formula unit (f.u.) above the G-AFM case with the highest magnetic moment ($\sim$ 5.0 $\mu_{\rm B}$ per Mn site). 

The benchmarking is extended in the comparison of the polar/centrosymmetric end members and ordered compounds at 25\% dopant molar concentration (cf. Fig. \autoref{fig:Magnetic-benchmarking}b). Here, polar (purple cross), centrosymmetric (teal triangle), and ordered compounds (orange plus) are indicated. Polar end members and their ordered compounds tend to have larger magnetic moments than those of the centrosymmetric end members, which is consistent with trends concerning the polar Mn-N bond lengths (see main text for discussion). Notably, the magnetic moments of ordered compounds remain centered around the magnetic moment of their polar end members with little deviation, even at 25\% dopant molar concentration, indicating the stability of the magnetic moment concerning cation substitution. Although the G-AFM ordering generally had the lowest magnetic moment for the polar end members and their ordered compounds ($\sim$ 4.50 $\mu_{\rm B}$ per Mn site), the lowest magnetic moment for the centrosymmetric end members was found in the A-AFM ordering ($\sim$ 3.75 $\mu_{\rm B}$ per Mn site), despite G-AFM ordering being the most stable for the centrosymmetric end members. 
Multiple factors influence existing trends between magnetic exchange and bond length (ionicity). While shorter bonds are correlated with stronger interactions \cite{Johannes2010}, the onset of increased bond covalency is found to be antagonistic to magnetic exchange \cite{Belashchenko2008}. For the centrosymmetric phases, larger Mn-N bond lengths are observed with a simultaneous increase in bond covalency, as evidenced by their metallic behavior, leading to a diminished magnetic moment.

\clearpage

 \section{Ferroelectricity  reversal}
\begin{figure}[ht]
    \centering
    \includegraphics[width=\linewidth]{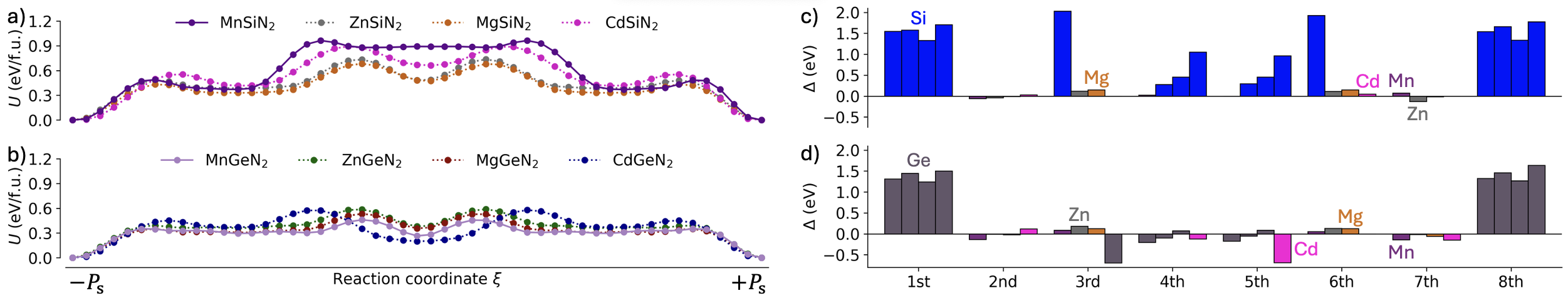}
    \caption{Minimum energy pathways of the (a) $A$SiN$_2$ and (b) $A$GeN$_2$ series computed \textit{via} nudged elastic band with the reaction coordinate $\xi$ from negative to positive polarity ($-P_{\rm s} \rightarrow +P_{\rm s}$). The MnSiN$_2$ and MnGeN$_2$ end members are shown as solid lines for clarity. Some end members exhibit large concavity in their respective minimum energy pathways. The atomic switching barrier $\Delta$ from the 1st to 8th metal-nitrogen pair for the (c) $A$SiN$_2$ and (d) $A$GeN$_2$ series. The series for the 1st to 8th cation is organized based on the legend order from the corresponding minimum energy pathways. The Mn (purple), Zn (gray), Mg (orange), Cd (pink), Ca (teal), Si (royal blue), and Ge (periwinkle) cations are labeled.}
    \label{fig:NEB-atomic-probe}
\end{figure}

\begin{figure}[ht]
    \centering
    \includegraphics[width=0.5\linewidth]{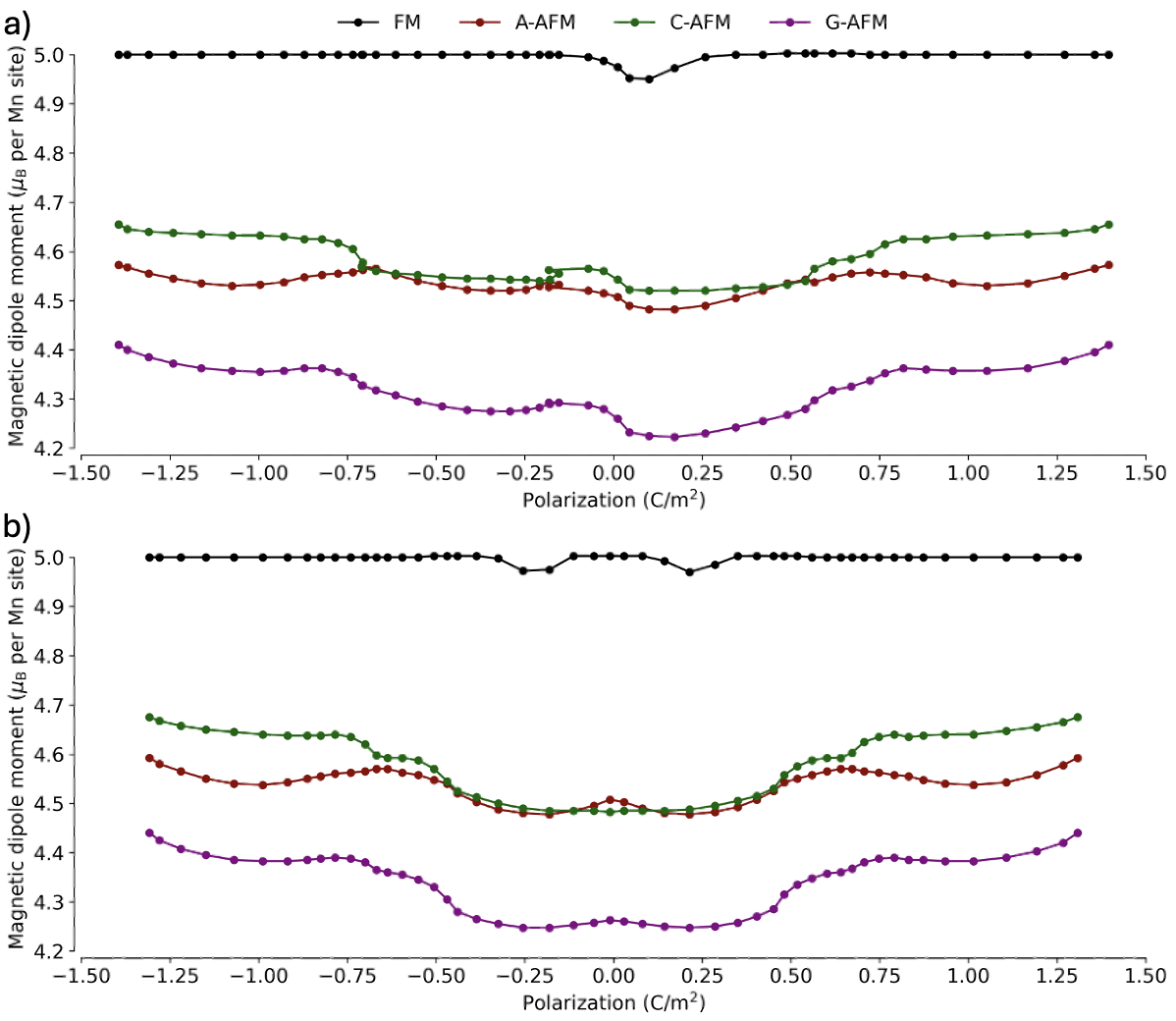}
    \caption{Magnetic dipole moment per Mn site as a function of the electric polarization in (a) MnSiN$_2$ and (b) MnGeN$_2$. The ferromagnetic (FM) and three antiferromagnetic (AFM) cases were considered. Here, the FM case is black, the A-AFM case is maroon, the C-AFM case is green, and the G-AFM case is purple. The magnetic moment remained stable throughout FE polarization reversal with G-AFM exhibiting the most significant deviation at 0.2 $\mu_{\rm B}$ per Mn site. The magnetic moment profile in MnSiN$_2$ is asymmetric, whereas the magnetic moment profile in MnGeN$_2$ is symmetric with shallow valleys near $\pm$ 0.25 C/m$^2$, likely due to the intermediate images of the reversal pathways.}
    \label{fig:Magnetization-versus-polarization}
\end{figure}

 The minimum energy pathways as a function of the reaction coordinate $\xi$ (from $-P_{\rm s}$ to $+P_{\rm s}$) of the end-member structures, in addition to the corresponding reversal barriers of each metal-nitrogen pair (\autoref{fig:NEB-atomic-probe}), illustrate key similarities and differences between the $A$SiN$_2$ and $A$GeN$_2$ series (see discussion in Sec. III(B) of the main text). 
 %
 During FE polarization reversal, the local magnetic moment did not significantly change for MnSiN$_2$ and MnGeN$_2$ (\autoref{fig:Magnetization-versus-polarization}). While no reversal of the magnetic moment is observed, the FM case showed the least perturbation to the magnetic moment at less than 0.1 $\mu_{\rm B}$ per Mn site, whereas the G-AFM case showed the largest (but still minor) perturbation at 0.2 $\mu_{\rm B}$ per Mn site. Notably, the magnetic moment profile of MnSiN$_2$ is asymmetric, while the magnetic profile of MnGeN$_2$ is symmetric. This observation can be rationalized by recognizing that the reversal pathways are distinct, with MnSiN$_2$ passing through a non-polar phase with significant distortions and MnGeN$_2$ passing through an anti-polar phase with minimal structural distortions. Interestingly, unlike MnSiN$_2$, there exist shallow valleys in the magnetic moment profile of MnGeN$_2$ at $\pm$ 0.25 C/m$^2$.
 
\begin{figure}[ht]
    \centering
    \includegraphics[width=0.7\linewidth]{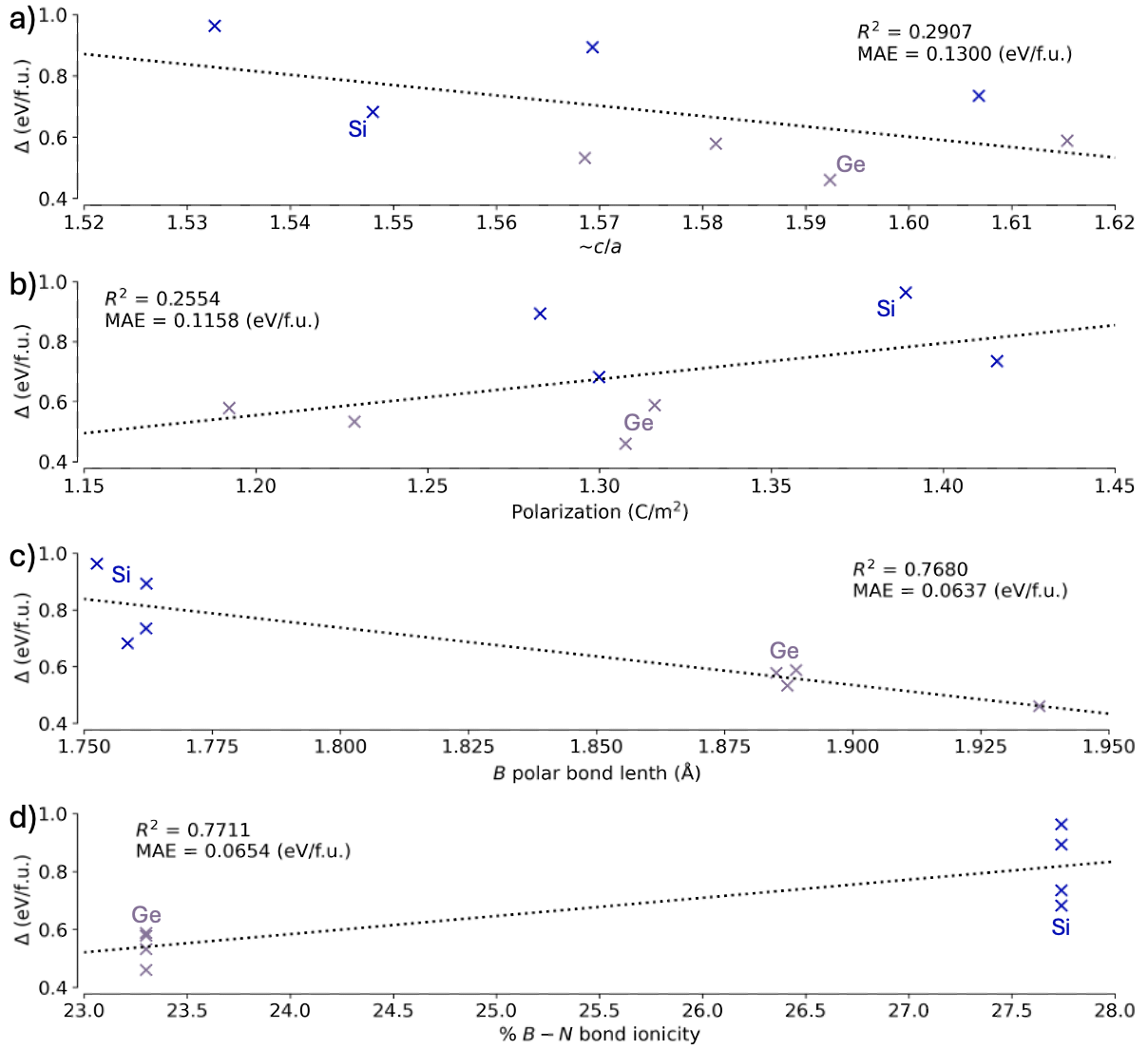}
    \caption{Benchmark of trends of the FE reversal barrier ($\Delta$) with respect to (a) $\sim c/a$, (b) electric polarization, (c) $B$ polar bond length, and (d) $B$-N bond ionicity. Silicon and germanium nitrides are denoted as blue and lilac, respectively. The $\sim c/a$ and $\Delta$ are negatively correlated with a coefficient of determination ($R^2$) of 0.2907 and mean absolute error (MAE) of 0.1300 eV/f.u., indicating a poor descriptor for $\Delta$. The electric polarization and $\Delta$ are positively correlated with an $R^2$ of 0.2554 and mean absolute error (MAE) of 0.1158 eV/f.u., indicating a poor descriptor for $\Delta$. The $B$ polar bond length and $\Delta$ are negatively correlated with an $R^2$ of 0.7680 and mean absolute error (MAE) of 0.0637 eV/f.u., indicating a great descriptor for $\Delta$. The $B$-N bond ionicity percentage and $\Delta$ are positively correlated with an $R^2$ of 0.7711 and mean absolute error (MAE) of 0.0654 eV/f.u., indicating a great descriptor for $\Delta$. It should be noted that only two data points ($i.e.,$ Si, Ge) are possible in this trend.}
    \label{fig:Reversal-benchmarking}
\end{figure}

 \autoref{fig:Reversal-benchmarking} summarizes the trends of the FE polarization reversal barrier ($\Delta$) with respect to important descriptors such as $\sim c/a$ (\autoref{fig:Reversal-benchmarking}a), electric polarization (\autoref{fig:Reversal-benchmarking}b), $B$ polar bond length (\autoref{fig:Reversal-benchmarking}c), and $B$-N bond ionicity (\autoref{fig:Reversal-benchmarking}d).  The conventional picture of the $c/a$ ratio is such that as this ratio approaches unity, $\Delta$ is expected to decrease since it approaches the centrosymmetric phase. However, the $\sim c/a$ and $\Delta$ are negatively correlated with a low coefficient of determination ($R^2$) of 0.2907 and a large mean absolute error (MAE) of 0.1300 eV/f.u., implying that large $c/a$ ratios decrease $\Delta$. Although the $c/a$ ratio has been proposed as a descriptor for ferroelectricity in wurtzites, recent evidence has indicated that there is no significant correlation between $\Delta$ and the $c/a$ ratio \cite{Fichtner2019, Lee2024-2, Fichtner2025}, consistent with our first-principles data. Likewise, the electric polarization and $\Delta$ are positively correlated with an $R^2$ of 0.2554 and mean absolute error (MAE) of 0.1158 eV/f.u., implying that the electric polarization is a poor descriptor of $\Delta$. 
 
 Considering local descriptors of ferroelectricity, the $B$ polar bond length and $\Delta$ are negatively correlated with an $R^2$ of 0.7680 and mean absolute error (MAE) of 0.0637 eV/f.u., which is a substantial improvement in the prediction of $\Delta$. This observation can be understood by recalling that long bonds generally are weaker, promoting bond breaking during FE polarization reversal. Similarly, the $B$-N bond ionicity percentage and $\Delta$ are positively correlated with an $R^2$ of 0.7711 and mean absolute error (MAE) of 0.0654 eV/f.u. Although this trend appears to contradict existing trends that bond ionicity decreases $\Delta$ \cite{Lee2024-2, Fichtner2025}, only two data points ($i.e.,$ Si, Ge) are possible in our dataset. Consistent with recent design recommendations of wurtzite FEs \cite{Lee2024-2}, local chemical descriptors ($e.g.$ bonding, ionicity) have substantively more accurate predictive power than conventional global predictors ($e.g.,$ $c/a$ ratio, electric polarization).

\clearpage
 \section{Magnetization reversal}
 
\begin{figure}[ht]
    \centering
    \includegraphics[scale=0.42]{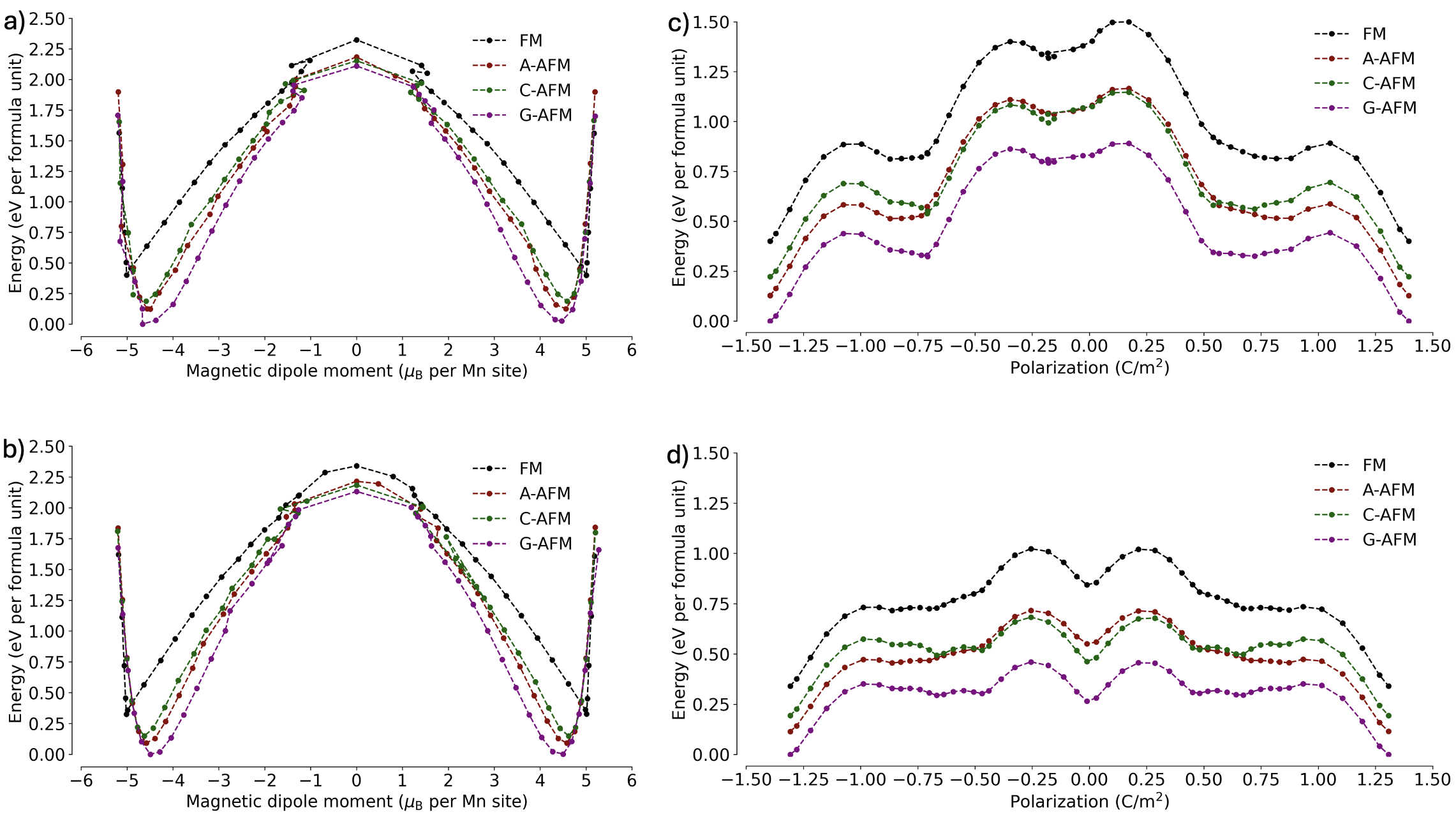}
    \caption{Magnetic reversal pathways per magnetic ordering at a fixed poled state for (a) MnSiN$_2$ and (b) MnGeN$_2$. The ferromagnetic (FM) and three antiferromagnetic (AFM) cases were considered. Here, the FM case is black, the A-AFM case is maroon, the C-AFM case is green, and the G-AFM case is purple. The pathways with G-AFM ordering had the lowest reversal barrier at approximately 2.00 eV per formula unit (f.u.), with the most stable polar states. The FE polarization reversal pathways per magnetic ordering of (a) MnSiN$_2$ and (b) MnGeN$_2$. The G-AFM ordering is indicated to have the lowest reversal barrier in both cases at approximately 0.8 and 0.5 eV per f.u. for MnSiN$_2$ and MnGeN$_2$, respectively. The G-AFM ordering case does not intersect with other magnetic ordering cases, indicating that these magnetic states are stable even under FE reversal.}
    \label{fig:FM-FE-reversal}
\end{figure}

 Magnetic reversal was simulated by constraining the magnetization of a fixed polar state to generate a distortion pathway as a function of the magnetic dipole moment per Mn site. The MnSiN$_2$ and MnGeN$_2$ end members with four magnetic ordering cases were considered here (cf. \autoref{fig:FM-FE-reversal}a,b). The G-AFM ordering had the smallest magnetic reversal barrier and most stable polar states at approximately 2.00 eV/f.u.\ in both cases. The FE polarization reversal pathway as a function of the magnetic ordering was compared between MnSiN$_2$ and MnGeN$_2$ (cf.\autoref{fig:FM-FE-reversal}c,d). The G-AFM ordering is indicated to have the lowest reversal barrier in both cases at approximately 0.8 and 0.5 eV/f.u.\ in MnSiN$_2$ and MnGeN$_2$, respectively. The distortion pathway does not deviate from the G-AFM ordering due to the lack of intersections, which implies significant stability of the G-AFM ordering even under FE polarization reversal. This observation is also evidenced by the stark difference in the magnetic and FE reversal barriers, which are factors of two and four in MnSiN$_2$ and MnGeN$_2$, respectively.

\newpage

\section{Ferroelectric polarization reversal and nonrelativistic spin splitting}

\begin{figure}[ht]
    \centering
    \includegraphics[width=0.5\linewidth]{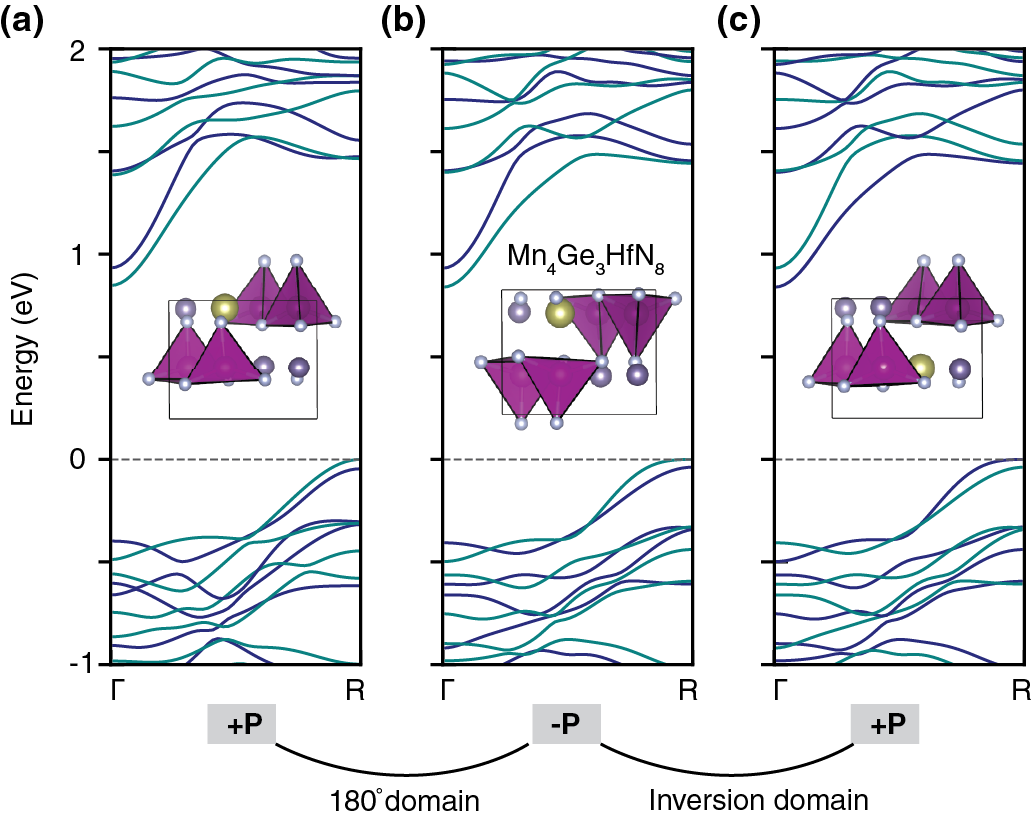}
    \caption{Nonrelativistic spin splitting (NRSS) in cation-ordered ferroelectric (a,b,c) Mn$_4$Ge$_3$HfN$_8$. Energy band dispersions of Mn$_4$Ge$_3$HfN$_8$in the (a) positively, (b) negatively, and (c) positively polarized state, respectively. (a) and (b) are mirror domains, while (a) and (c) are inversion domains. Signs of NRSS are reversed between the inversion domains but not the mirror domains.}
    \label{fig:Mn4X3YN8_nrss_bands}
\end{figure}

\autoref{fig:Mn4X3YN8_nrss_bands} presents the spin-polarized DFT--PBE band structure for the $-P$ and $+P$ electric polarization configurations for cation-ordered Mn$_4$Ge$_3$HfN$_8$, without relativistic spin-orbit coupling. 
The insulating band gap is $\sim$0.8\,eV. A sizable nonrelativistic spin splitting (NRSS) appears at the valence and conduction band edges along the low-symmetry path from $\Gamma$(0,0,0) to R($\pi$,$\pi$,$\pi$), reaching up to 92 and 238 meV, respectively. 
Unlike their end members (MnGeN$_2$ and MnHfN$_2$), NRSS is also observed at $\Gamma$, because of the additional symmetry breaking introduced by ordered cation substitution \cite{Yuan2024}. 
\autoref{fig:Mn4X3YN8_nrss_bands}a,c exhibit $+P$ electric polarizations, but they differ in their relation to the $-P$ configuration (\autoref{fig:Mn4X3YN8_nrss_bands}b). While (a) and (b) correspond to 180$^\circ$ mirror twin domains (they are related by mirror $m_z$), (b) and (c) are inversion domains (they are related by inversion). 
Comparing the electronic band structure for all three configurations, we find polarization-driven reversal of NRSS occurs only for the inversion domain pair [\autoref{fig:Mn4X3YN8_nrss_bands}(b,c)]. 
These cation-ordered variants together with their end members constitute a compelling material space for designing electric-field-switchable multiferroic devices \cite{PhysRevLett.134.106802,urru2025g}.

\pagebreak
\newpage
\bibliography{references}